\documentclass[eqseqnum,twocolumn]{revtex4}
\usepackage{amsmath,amsthm,amsfonts,amssymb}
\usepackage{graphicx}
\usepackage{subfigure}
\usepackage[usenames]{color}

\renewcommand{\thesubfigure}{\thefigure.\arabic{subfigure}} 
\makeatletter 
\renewcommand{\@thesubfigure}{(\thesubfigure)\space} 
\renewcommand{\p@subfigure}{} 
\makeatother 
\numberwithin{equation}{section}
\renewcommand{\vec}[1]{{\mathbf #1}}        

\newcommand{\ten}[1]{{\mathbf{#1}}{}}
\newcommand{\Id}{\ten{I}}
\newcommand{\thetabold}{\mbox{\boldmath{$\theta$}}}

\begin{document}
\title{The interplay between boundary conditions and flow geometries in
    shear banding: hysteresis, band configurations, and surface
    transitions}
\date{\today}
\author{J. M. Adams}
\affiliation{Cavendish Laboratory, University of Cambridge, JJ Thomson
   Avenue,Cambridge CB3 0HE, U.K.}
\author{S. M. Fielding}
\affiliation{School of Mathematics, University of Manchester, Booth
   Street East, Manchester, M13 9EP, U.K.}
\author{P. D. Olmsted}
\affiliation{School of Physics \& Astronomy, University of Leeds,
   Leeds, LS2 9JT, U.K.}
\begin{abstract}
  We study shear banding flows in models of wormlike micelles or
  polymer solutions, and explore the effects of different boundary
  conditions for the viscoelastic stress. These  are needed because the
  equations of motion are inherently non-local and include
  ``diffusive'' or square-gradient terms. Using the diffusive
  Johnson-Segalman model and a variant of the Rolie-Poly model for
  entangled micelles or polymer solutions, we study the interplay
  between different boundary conditions and the intrinsic stress
  gradient imposed by the flow geometry. We consider prescribed
  gradient (Neumann) or value (Dirichlet) of the viscoelastic stress
  tensor at the boundary, as well as mixed boundary conditions in
  which an anchoring strength competes with the gradient contribution
  to the stress dynamics. We find that hysteresis during shear rate
  sweeps is suppressed if the boundary conditions favor the state
  that is induced by the sweep. For example, if the boundaries favor
  the high shear rate phase then hysteresis is suppressed at the low
  shear rate edges of the stress plateau. If the boundaries favor the
  low shear rate state, then the high shear rate band can lie in the
  center of the flow cell, leading to a three-band configuration.
  Sufficiently strong stress gradients due to curved flow geometries,
  such as that of cylindrical Couette flow, can convert this to a
  two-band state by forcing the high shear rate phase against the wall
  of higher stress, and can suppress the hysteresis loop observed
  during a shear rate sweep.  

\end{abstract}
\maketitle
\section{Introduction}
\begin{figure}[!htb]
\begin{center}
\includegraphics[width = 0.48\textwidth]{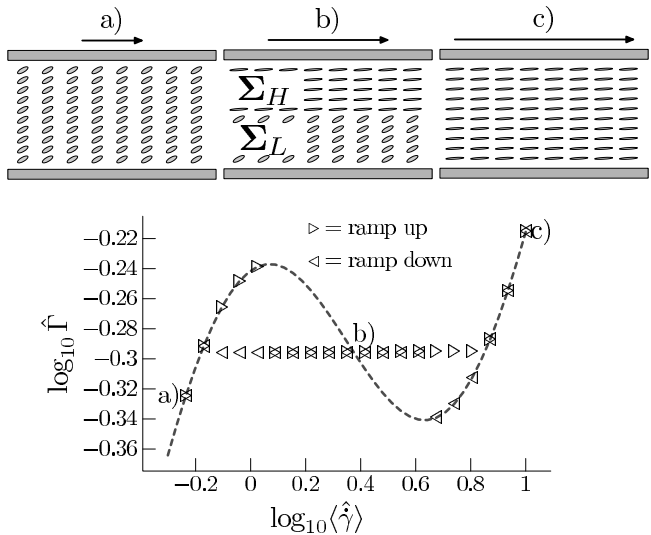}
\end{center}
\caption{Dashed line: the constitutive curve (dimensionless specific
  torque $\hat{\Gamma}$ as a function of dimensionless shear rate
  $\hat{\dot{\gamma}}$) for the diffusive Johnson-Segalman (DJS) model
  in cylindrical Couette flow. The triangles show simulated steady
  state flow curves for zero gradient (Neumann) boundary conditions,
  for increasing ($\triangleright$) and decreasing ($\triangleleft$)
  the average shear rate $\langle\hat{\dot{\gamma}}\rangle$ from small
  or large average shear rates respectively.  The model parameters are
  $a=0.3$, $\epsilon = 0.05$, $\hat{\mathcal{D}} = 10^{-5}$, and the
  gap size is 0.1\% of the radius ($q = 0.005$).  The physical
  difference between the two bands is illustrated by ellipsoids whose
  principal axes' directions and length ratios coincide with those of
  the viscoelastic stress tensor $\ten{\Sigma}_L$ and $\ten{\Sigma}_H$
  in, respectively, the low and high shear rate phases.}
\label{fig:djsneumann}
\end{figure}

Microscopic models for certain viscoelastic fluids, such as wormlike
micelles \cite{Cate96} and high molecular weight polymeric liquids
\cite{doiedwards,MilnerML01} predict unstable homogeneous stationary
states for controlled average shear rates. Specifically, the coupled
equations of motion for the fluid flow and microstructural quantities,
such as the viscoelastic stress $\ten{\Sigma}$ or orientational order,
predict a nonmonotonic \textit{constitutive curve}, \textit{i.e.} the
steady state total shear stress $T_{xy}$ (or equivalently the applied
torque) as a function of applied shear rate $\dot{\gamma}$ for
homogeneous flows.  This constitutive curve typically is multivalued
with a region of decreasing shear stress as a function of increasing
shear rate, which is a hallmark of hydrodynamic instability
\cite{renardybook}.  Understanding this nonlinear behavior is
important for practical applications such as injection moulding of
plastics and drilling muds used in bore holes \cite{Son1973}. The
simplest resolution of the instability is an inhomogeneous state with
macroscopic regions of high and low shear rates, known as shear bands
\cite{olmsted92,spenley96}, which has been widely observed in wormlike
micelles \cite{schmitt94,BritCall97c,Berr97,GAC97}, lamellar
surfactant solutions \cite{Salmon.Manneville.ea03}, and liquid
crystals \cite{pujolle01}. The resulting experimental signature of
this inhomogeneous state is a stress plateau as  a function of
controlled average shear rate. We denote  this
experimentally determined relation between total shear stress and
average shear rate as the \textit{flow curve}; in shear banding flow
this curve incorporates inhomogeneous flows and is distinct from the
constitutive curve, which cannot be measured because of the
constitutive instability. In Fig.~\ref{fig:djsneumann} the dashed line
shows the constitutive curve while the triangles show flow curves that
could be measured upon either increasing or decreasing average shear
rate ramps.  

Experimentally, it is clear that the geometry of the shear cell
affects the positioning of the shear bands.  The simplest structure is
seen in the cylindrical Couette geometry where two bands form, with
the higher shear rate band near the inner wall (rotor)
\cite{Salmon.Colin.ea03}.  In the cone and plate geometry a different
band configuration has been reported: two low shear rate regions next
to the cone and plate separated by a high shear rate region
\cite{BritCall97c}. One explanation offered for that result was that
secondary flows stabilized the center band \cite{KumarL00}. We will
show that a boundary condition that prescribes a value for the
polymeric stress tensor similar to that of the low shear rate phase
can induce this three-band configuration.

The experimental rheological features of shear banding have been
studied in detail for wormlike micellar solutions
\cite{Berr97,GAC97,lerouge00}. A constitutive curve such as the dashed
line in Fig.~\ref{fig:djsneumann} can only be inferred because the
negative slope region is unstable; however, the experimentally
measured flow curve, indicated by triangles, typically shows a stress
plateau which spans the range of average applied shear rates at which
the system shows shear banding. During shear rate ramps from rest,
hysteresis is often observed at the start of this plateau: the stress
increases past the plateau stress and follows the constitutive curve
until, at a time that depends on the rate of the ramp, a high shear
rate band develops and the total stress decreases onto the steady
state plateau \cite{Berr97,GAC97,BerrPort99}. However, if the shear
rate is reduced from a point on the plateau, the low shear rate branch
of the constitutive curve is intersected directly by the flat plateau.
This behavior is reminiscent of nucleation and hysteresis at first
order phase transitions, and we will show below that the boundary
conditions can influence this hysteresis, in a manner analogous to
heterogeneous nucleation.

In modelling shear banding systems such as wormlike micelles the most
important degrees of freedom are the fluid velocity $\vec{v}$ and a
viscoelastic contribution to the stress, $\ten{\Sigma}$; here we
neglect other other potentially important quantities such as
concentration and micellar length.  Among models with non-monotonic
constitutive curves for homogeneous flow, spatially local ones such as
the Johnson-Segalman model in its original form \cite{johnson77} show
high sensitivity to the flow history and do not give the well-defined
and unique stress plateau observed in experiments \cite{GrecBall97}.
In contrast, if a non-local diffusive term (typically proportional to
$\nabla^2\ten{\Sigma}$) is included in the constitutive equation for
the viscoelastic stress (\textit{e.g.}  the diffusive Johnson-Segalman
(DJS) model), there is only a single value of the total stress for
which the interface between bands is stable and stationary
\cite{olmsted99a}.  The diffusive term arises from microscopic
physical mechanisms such as the diffusion of molecules that carry
stress \cite{elkareh89}, the persistence length of wormlike micelles
\cite{liu93}, or hydrodynamics \cite{Dhon99,LOB00}. The resulting
equation governing inhomogeneous steady states is thus a spatial
differential equation for the viscoelastic stress, which necessitates
a boundary condition that is, at present, unknown.  The most frequently
used boundary condition on the viscoelastic stress $\ten{\Sigma}$ has
been that of zero stress gradient parallel to the boundary normal
\cite{olmsted99a}. This boundary condition predicts the flow curve
shown in Fig.~\ref{fig:djsneumann} for the DJS model. As noted above,
this flow curve shows hysteretic behavior at the start of the stress
plateau similar to that seen in experiment \cite{GAC97}, with a stress
overshoot during an increasing shear rate ramp.

Other theoretical studies of shear banding have used a fixed value of
the polymeric stress
\cite{Cook.Rossi04,rossimckinley06,PicAjdBocLeq02} at the wall
(Dirichlet boundary conditions). Cook and Rossi considered a two-fluid
model for wormlike micelles in which the micelles were assumed to
align at the wall parallel to the flow direction, in both planar and
cylindrical Couette flow. They found that the flow curve deviates
significantly from the constitutive curve near the low shear rate
branch \cite{Cook.Rossi04,rossimckinley06}, exhibited hysteresis only
at the high shear rate side of the stress plateau, and possessed a
three-band state with the boundary condition induced high shear rate
bands near the wall.  Qualitatively similar results were found by
Picard and co-workers in a scalar model for shear banding in a yield
stress solid \cite{PicAjdBocLeq02}.


In this paper we perform a more detailed study of the effects of
different boundary conditions for the viscoelastic stress at the wall,
$\ten{\Sigma}_0$. We consider strong ``anchoring'' in which Dirichlet
conditions apply, and interpolate between oblate and prolate forms of
$\ten{\Sigma}_0$. We also study mixed boundary conditions, in which
$\ten{\Sigma}_0$ is determined by a balance between spatial gradients and
surface anchoring. Moreover, we address the important
interplay between the boundary conditions and intrinsic inhomogeneity
of the flow determined by the flow geometry (to compare cylindrical
and planar Couette flow with cone and plate flow, for example).

The organization of the paper is as follows. In Section
\ref{sec:equations-motion} we outline the details for solving the
equations of motion within the creeping flow approximation in
cylindrical Couette flow, and describe two constitutive equations for
the viscoelastic stress: the DJS model \cite{olmsted99a} and a
tube-based model which has been developed for polymer solutions and
wormlike micelles \cite{likhtmangraham03,challenges96}. The physical
motivation for the boundary conditions is discussed in Section
\ref{sec:bc}. After reviewing previous results for Neumann boundary
conditions in Section \ref{sec:neumann}, the main new results are
presented in Section \ref{sec:results}. We organize these as follows:
(1) the effect of different fixed value (Dirichlet) boundary
conditions on the flow curves and hysteresis upon increasing and then
decreasing the average shear rate; (2) the interplay between boundary
conditions and the stress gradient imposed by the flow geometry
(\textit{e.g.} cylindrical Couette flow) and their effects on the
stable position of the shear bands; and (3) the effect of mixed
boundary conditions, in which the viscoelastic stress at the surface
is influenced by both the bulk constitutive relation and the flow
geometry. We demonstrate the possibility of a transition between
different effective boundary conditions as a function of the anchoring
strength.
\section{Equations of Motion}\label{sec:equations-motion}
\subsection{Creeping Flow Approximation}\label{sec:creep-flow-appr}
In a wormlike micellar system we assume that the total stress
$\ten{T}$ can be separated into contributions from the Newtonian
solvent and a viscoelastic stress $\ten{\Sigma}$ from the micelles:
\begin{equation}
\ten{T} =   2 \eta \ten{D} + \ten{\Sigma} -p \Id.
\label{eqn:totalstress}
\end{equation}
Here $\Id$ denotes the identity matrix, $\eta$ is the solvent shear
viscosity, $\ten{D} = \tfrac12\left[\nabla \vec{v} + (\nabla
  \vec{v})^T\right]$ is the symmetric velocity gradient tensor, and
$p$ is the isotropic pressure determined by incompressibility ($\nabla
\cdot \vec{v} = 0$). We are interested here in the creeping flow
regime (low Reynolds number), for which the force balance has the form
\begin{equation}
\nabla\cdot \ten{T}=0.
\label{eqn:eom}
\end{equation}
We perform calculations explicitly for Couette flow between concentric
cylinders and assume unidirectional flow $\vec{v}=v(r,t)
\hat{\vec{\thetabold}}$, where $r$ and $\theta$ are the usual
cylindrical coordinates.  In this case,
Eqs.~(\ref{eqn:totalstress}-\ref{eqn:eom}) imply that
\begin{equation}
\eta \dot{\gamma}(r,t) + \Sigma_{r\theta}(r,t) = \frac{\Gamma}{r^2},
\end{equation}
where 
\begin{equation}
\dot{\gamma}(r,t) = r \frac{\partial}{\partial r}\frac{v(r,t)}{r},
\end{equation}
and $\Gamma$, the specific torque per unit cylinder height per radian,
is a constant of integration. We assume no-slip boundary conditions on
the velocity field. For cylindrical Couette flow
with a fixed outer cylinder and a rotating inner cylinder of velocity
$V$, the no-slip boundary conditions lead to the following global
constraint
\begin{equation}
\frac{V}{R_1} = \int_{R_2}^{R_1} \dot{\gamma}(r,t) \frac{dr}{r},
\end{equation}
which implies that the specific torque $\Gamma$ is given by
\begin{equation}
\Gamma \frac{R_1^2 -R_2^2}{R_1^2 R_2^2} = 2 \left(\frac{\eta V}{R_1}-
\int_{R_1}^{R_2}\Sigma_{r \theta}\frac{dr}{r} \right).
\end{equation}

To simplify the equations we define the spatial variable
\begin{equation}
x = \frac{1}{q} \ln \left( \frac{r}{R_1}\right),
\end{equation}
where $q = \ln \frac{R_2}{R_1}$. Making use of this variable change,
the creeping flow equation becomes
\begin{equation}
\eta \dot{\gamma} = \Gamma e^{-2 q x} - \Sigma_{r \theta},
\label{eqn:gammasub}
\end{equation}
with  the specific torque is given by
\begin{equation}
\Gamma = \frac{2q}{1-e^{-2 q}}\left( \langle\Sigma_{r \theta}\rangle - 
\eta \langle \dot{\gamma}\rangle\right),\label{eq:torque}
\end{equation}
where the spatial averages are simplified to $\langle(\cdot)\rangle =
\int_0^1 (\cdot) dx$. This expression can be used to calculate
$\Gamma$ for an imposed average shear rate, and hence render
Eq.~(\ref{eqn:gammasub}) an integral equation determining the relation
between the local and average shear rates and viscoelastic shear
stress.  A complete description of the dynamics requires an equation
of motion for the viscoelastic stress $\ten{\Sigma}$,
examples of which are described below.
\subsection{Constitutive Equations}
\label{sec:constitutive}
\subsubsection{Diffusive Johnson-Segalman (DJS) Model}
One of the simplest tensorial models that produces a non-monotonic
constitutive curve is the DJS model, which has been studied in detail
 \cite{johnson77,malkus90,GrecBall97,olmsted99a}.
Physically it is motivated by neglecting all but the lowest modes of
vibration of polymer chains, thereby representing them as elastic
dumbbells with span $\vec{R}$ \cite{phanthien02}. The viscoelastic
stress tensor $\ten{\Sigma}$ is related to the extension of the
dumbbells by $\ten{\Sigma} = G(-\Id + k \langle \vec{R}\vec{R}
\rangle)$, where expressions for the normalization $k$ and the modulus
$G$ can be derived in terms of the parameters in the underlying
microscopic polymeric models. Here $\langle(\cdot)\rangle$ denotes a
thermal average. The DJS model \cite{larson88,olmsted99a} provides the
constitutive equation for the evolution of the viscoelastic stress:
\begin{align}
  \stackrel{\blacklozenge}{\ten{\Sigma}}+\frac{1}{\tau} \ten{\Sigma} &= 
  2 \frac{\mu}{\tau}\ten{D} +{\mathcal D} \nabla^2
  \ten{\Sigma},&&
  \label{eqn:jsmodel}
\end{align}
where
\begin{equation}
\stackrel{\blacklozenge}{\ten{\Sigma}} = (\partial_t + \vec{v}\cdot \nabla)
\ten{\Sigma}+(\ten{\Omega}\cdot \ten{\Sigma} - \ten{\Sigma}\cdot 
\ten{\Omega})-a(\ten{D} \cdot\ten{\Sigma} + \ten{\Sigma}\cdot \ten{D}),
\end{equation}
$\tau$ is a relaxation time, the ``polymer'' viscosity $\mu$ is
related to the modulus by $G=\mu/\tau$, and $\ten{\Omega}
=\tfrac12\left[ \nabla \vec{v} - (\nabla \vec{v})^T \right]$. The
total stress comprises the viscoelastic stress of the DJS model and a
Newtonian contribution, according to Eq.~(\ref{eqn:totalstress}), and
the viscosity ratio $\epsilon\equiv\eta/\mu$ controls the balance
between the two stresses. The `slip parameter' $a$, which describes
non-affine stretch of the dumbell with respect to the extension of the
flow, allows for a non-monotonic constitutive curve for $0<|a|<1$ and
$\epsilon<1/8$.

The ``diffusion'' term $\mathcal D\nabla^2\ten{\Sigma}$ describes
non-local relaxation of the viscoelastic stress and is necessary to
describe strongly inhomogeneous flow profiles \cite{olmsted99a}.
Because of this term the steady shear banding state obeys a spatial
differential equation, which must solved subject to boundary
conditions specified at the walls of the flow cell. The solvability
condition for a stationary interface leads to a unique total shear
stress plateau for imposed average shear rates in the non-monotonic
portion of the constitutive curve \cite{LOB00}.  The characteristic
width $\ell$ of the interface between shear bands is given by
$\ell=\sqrt{{\mathcal D}\tau}$.

For convenience we define the modulus $G=\mu/\tau$. In cylindrical
Couette flow the constitutive equation has the following components
\cite{olmsted99a}:
\begin{subequations}
\label{eq:DJS}
\begin{eqnarray}
  {\mathcal L }\,\Sigma_{rr} &=& - (1-a)\dot{\gamma}\Sigma_{r\theta} +
  \frac{2{\mathcal D }e^{-2qx}}{q^2R_1^2}\left(\Sigma_{\theta\theta} -
  \Sigma_{rr}\right) \\ 
  {\mathcal L }\,\Sigma_{\theta\theta} &=&
  \phantom{-}(1+a)\dot{\gamma}\Sigma_{r\theta} -
  \frac{2{\mathcal D
  }e^{-2qx}}{q^2R_1^2}\left(\Sigma_{\theta\theta}-\Sigma_{rr}\right) \\ 
\nonumber
  {\mathcal L }\,\Sigma_{r\theta} &=&  \dot{\gamma}\left[G-
    \frac{1-a}{2}\Sigma_{\theta\theta} +
    \frac{1+a}{2}\Sigma_{rr}\right] - \frac{4
      {\mathcal D}e^{-2qx}}{q^2R_1^2}\Sigma_{r\theta},\\
\end{eqnarray}
\end{subequations}
where
\begin{equation}
  {\mathcal L } \equiv \frac{\partial}{\partial t} + \frac{1}{\tau} -
  \frac{{\mathcal D   }e^{-2qx}}{q^2 R_1^2}\frac{\partial^2}{\partial x^2}.
\end{equation}

\subsubsection{Reptation-Reaction and Rolie-Poly Models}
In Cates' model \cite{cates90} of wormlike micellar solutions the
micelles are assumed to relax in a tube, and the deviatoric
viscoelastic stress is defined by $\ten{\Sigma} = G(-\Id+3 \langle
\vec{u} \vec{u} \rangle)$, where the unit vector $\vec{u}$ describes
the local orientation of tube segments. The viscoelastic stress is
given by a history integral over the second moment of the correlation
function $\langle \vec{u} \vec{u}\rangle$, weighted by the
distribution of tube segments as they are created and subsequently
vacated by the wormlike micelles. The resulting integral equation can
be re-written as the following differential equation by using a decoupling
approximation to remove fourth order moments \cite{challenges96}:
\begin{equation}
\stackrel{\nabla}{\ten{\Sigma}} + \frac{1}{\tau}\ten{\Sigma} = 
2 G \ten{D} - \frac{2}{3} \ten{\Sigma}:\nabla\vec{v} \left( \Id + (1+\beta)
[\ten{\Sigma}/G] \right) + {\mathcal D} \nabla^2 \ten{\Sigma},
\label{eqn:catesmodel}
\end{equation}
where
\begin{equation}
\stackrel{\nabla}{\ten{\Sigma}} = (\partial_t + \vec{v}\cdot \nabla)
\ten{\Sigma}+(\ten{\Omega}\cdot \ten{\Sigma} - \ten{\Sigma}\cdot 
\ten{\Omega})-(\ten{D} \cdot\ten{\Sigma} + \ten{\Sigma}\cdot \ten{D}).
\end{equation}
and $\beta = 0$.  A stress diffusion term, which did not appear in the
original formulation, has been included here.  The non-linear term in
this equation preserves the traceless property of the deviatoric
stress.

Interestingly, for $\beta\neq0$ this model is identical to an
extension of the original Doi-Edwards (DE) theory of entangled
polymers \cite{doiedwards}, which has a constitutive instability and
therefore predicts shear banding. This extension to the DE theory
incorporates the enhanced release of polymer entanglements due to
convection\cite{Marr96b}, which increases the polymer stress and can
``cure'' the DE instability for sufficiently strong convected
constraint release (CCR). CCR and tube stretching were incorporated in
\cite{MilnerML01} to describe polymer melts and wormlike micelles, and
later simplified to a differential version called the Rolie-Poly model
\cite{likhtmangraham03}.  In the version of the Rolie-Poly model used
in Eq.~(\ref{eqn:catesmodel}) the tube length is assumed to be relaxed
(``non-stretching'').  The CCR parameter $\beta$ ($0\leq\beta\leq1$)
corresponds to the frequency of the release of polymer entanglement
constraints due to convection by the flow, and
Ref.~\cite{likhtmangraham03} used $\beta=1$ to model a 
well-entangled polymer melt without a constitutive instability. For
small enough $\beta$ a constitutive instability results.

\begin{figure}[!htb]
\begin{center}
\includegraphics[width = 0.48\textwidth]{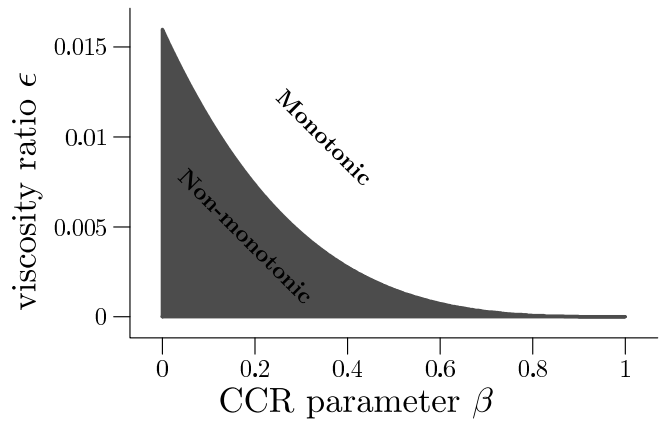}
\end{center}
\caption{Regions of parameter space $\beta$ and $\epsilon$ for which
  the DRP model has either a monotonic (unshaded) or
  non-monotonic (shaded) constitutive curve.}
\label{fig:betaepsilon}
\end{figure}

The DE constitutive instability is thought to be an unphysical
prediction of polymer melts \cite{MilnerML01}, but a series of recent
experiments have reignited the interest in this instability by showing
data, such as velocity profiles, consistent with shear banding in
polymer \textit{solutions}
\cite{Tapadia2006Direct-visualiz,Tapadia2006Banding-in-enta,hu2007cre}.
Hence, the Rolie-Poly model is an alternative simple constitutive
model to the DJS model for wormlike micelles, and by tuning the
parameter $\beta$ it can be used to address polymer solutions and melts.

As with the DJS model, we define the modulus by $G=\mu/ \tau$ and the
ratio of Newtonian (or ``solvent'') to viscoelastic stress by
$\epsilon=\eta/\mu$. For a given $\epsilon$ a non-monotonic
constitutive curve is found if the degree of convected constraint release
$\beta$ is small enough (Fig.~\ref{fig:betaepsilon}). We will refer to
the model described above as the diffusive Rolie-Poly (DRP) model.  In
cylindrical Couette flow the components of the DRP model evolve
according to
\begin{widetext}
\begin{subequations}
\label{eq:DRP}
\begin{eqnarray}
  {\mathcal L }\,\Sigma_{rr} &=&\!\! -\frac{2}{3}\dot{\gamma}\Sigma_{r\theta}
  \left[1 + (1+\beta)(\Sigma_{rr}/G)\right] + 
  \frac{2{\mathcal D }e^{-2qx}}{q^2R_1^2}\left(\Sigma_{\theta\theta} -
  \Sigma_{rr}\right) \\ 
  {\mathcal L }\,\Sigma_{\theta\theta} &=&\!\! -
  \frac{2}{3}\dot{\gamma}\Sigma_{r\theta} 
  \left[1 + (1+\beta)(\Sigma_{\theta\theta}/G)\right]
  + 2\dot{\gamma}\Sigma_{r\theta} -
  \frac{2{\mathcal D }e^{-2qx}}{q^2R_1^2}\left(\Sigma_{\theta\theta} -
  \Sigma_{rr}\right) \\ 
  {\mathcal L }\,\Sigma_{r\theta} &=&\!\! G\dot{\gamma} + \dot{\gamma}
       \Sigma_{rr} -
  \frac{2}{3}\dot{\gamma}\Sigma_{r\theta} 
   (1+\beta)[\Sigma_{r\theta}/G] + \frac{4
      {\mathcal D}e^{-2qx}}{q^2 R_1^2}\Sigma_{r\theta} .
\end{eqnarray}
\end{subequations}

\section{Boundary conditions}
\label{sec:bc}
\subsection{Liquid crystal analogy}
To motivate the possible boundary conditions in micellar systems, we
first recall the situation of nematic liquid crystals confined between
walls, in which the boundary conditions are a balance between surface
and bulk interactions \cite{pgdg}. The free energy depends on the
nematic order parameter $\ten{Q} = \langle \vec{u}\vec{u} \rangle
-\textstyle{\frac{1}{3}} \ten{I}$, where the unit vector $\vec{u}$ is
parallel to the liquid crystalline mesogen \cite{pgdg}. A simple free
energy functional for a nematic liquid crystal is \cite{pgdg}
\begin{equation}
  F_{\textrm{LC}} = \int_{V}\left[ f^b_{h}(\ten{Q})+\tfrac12 {\mathcal
    K} (\nabla \ten{Q})^2 \right]dV  
  +  \tfrac12 W_{\scriptscriptstyle\textrm Q} \int_{\textrm{walls}}\!\!\!\!\! {\rm Tr}
    (\ten{Q}-\ten{Q}_0)^2 dS, 
\end{equation}
\end{widetext}
where $f^b_{h}(\ten{Q})$ is the homogeneous part of the bulk free
energy, typically expanded in Landau form as
$f^b_{h}(\ten{Q})=\tfrac{a}{2} \textrm{Tr}\ten{Q}^2 + \ldots$; and
${\mathcal K}$ is a Frank elastic constant. In the surface free
energy, $W_{\scriptscriptstyle\textrm Q}$ specifies the strength of the wall potential which
penalizes deviations away from a specific order parameter tensor
$\ten{Q}_0$.  More complex surface free energies are possible
\cite{rey04}, and one may alternatively consider boundary conditions
that influence the orientation, but not magnitude, of $\ten{Q}$ at the
wall.

The spatial dependence of the order parameter is found by demanding
that the free energy functional be stationary with respect to varying
$\ten{Q}$. Stationarity in the bulk leads to the following
differential equation,
\begin{align}
  \label{eq:QDE}
  \frac{\delta F_{\textrm LC}}{\delta\ten{Q}} &= \left[ a Q +
  \ldots\right]  - {\mathcal K}\nabla^2Q=0,
 \end{align}
 whose boundary condition arises from requiring zero variation at the
 wall:
\begin{equation}
{\mathcal K}\,\boldsymbol{\hat{\textbf{n}}}\cdot \nabla\ten{Q}+ W_{\scriptscriptstyle\textrm Q}
(\ten{Q}-\ten{Q}_0) = 0.
\end{equation}
Here $\boldsymbol{\hat{\textbf{n}}}$ is the outward unit normal from
the fluid at the wall. Physically the boundary condition is a balance
of surface and bulk mesogen torques at the wall.

The bulk equation defines a correlation length $\ell=\sqrt{{\mathcal
    K}/a}$, which controls the decay of order parameter fluctuations
away from the surface, while the boundary defines an
\textit{extrapolation}
length 
\begin{equation}
  \label{eq:xiLC}
\xi = \frac{{\mathcal K}}{W_{\scriptscriptstyle\textrm Q}}.
\end{equation}
which controls wall anchoring \cite{pgdg}. The extrapolation length
$\xi$ roughly sets the scale for the gradient
$\nabla\simeq1/\xi$ induced at the wall by the boundary conditions, in
the absence of bulk Frank stresses. For small $\xi$ the effective
gradient at the wall is very large, so the wall potential has a
strong effect. Conversely, for large $\xi$ the characteristic gradient
is small, and the anchoring potential has a weak effect.

The extrapolation length should thus be compared with the bulk healing
length or correlation length $\ell$ to assess the relevant regime.
For strong anchoring ($\xi \ll \ell$) the boundary conditions
effectively impose $\ten{Q}_0$ at the walls, while for weak anchoring
($\xi \gg \ell$) the boundary conditions are dictated by the Frank
elastic term, resulting in Neumann boundary conditions, or
$\nabla\ten{Q}=0$.  Mixed boundary conditions apply between these two
extremes \cite{pgdg}.
\subsection{Viscoelastic stress}
Based on the liquid crystalline example, we apply similar boundary
conditions to the viscoelastic stress; the details are given in
Appendix \ref{sec:liqu-cryst-anal}. The viscoelastic stress is
analogous to the liquid crystalline order parameter, and the stress
diffusion constant is analogous to the Frank elastic constant
(according to ${\mathcal D}\tau \rightarrow {\mathcal K}$), and there
is a corresponding anchoring potential, leading to
\begin{equation}
{\mathcal D}\tau\,\boldsymbol{\hat{\textbf{n}}}\cdot \nabla\ten{\Sigma}+ W
(\ten{\Sigma}-\ten{\Sigma}_0) = 0. \label{eq:mixed}
\end{equation}
In this formulation the anchoring potential has the dimensions of
length.

As shown in Appendix \ref{sec:liqu-cryst-anal}, the anchoring
potential $W$ can be expressed as $W=W_0/G$, where $G$ is the modulus
and $W_0$, with dimensions of energy per area, penalizes deviations of
the polymer deformation $\ten{\Lambda}$ from a reference strain
$\ten{\Lambda}_0$. A simple, albeit weak, contribution to the
anchoring potential comes from the effect of the wall on nearby
micellar or polymer conformations. In Appendix
~\ref{sec:liqu-cryst-anal} we estimate $W\simeq 2R_g$, where $R_g$ is
the radius of gyration.

By analogy with the liquid crystal example above, the extrapolation
length is given by
\begin{align}
  \xi &= \frac{{\mathcal D}\tau}{W}=
  \frac{\sqrt{{\mathcal D}\tau}}{W}\,\ell,
\end{align}
where the interfacial width $\ell$ is set by the diffusion constant
according to $\ell=\sqrt{{\mathcal D}\tau}$. The two characteristic
lengths are comparable when $\xi_c/\ell\simeq1$, or as noted above, we
expect weak anchoring (large $\xi$) to yield effectively Neumann
boundary conditions (zero gradient)
\cite{olmsted99a,GrecBall97,rossimckinley06}.  Alternatively, very
strong anchoring might be encouraged by specific wall treatments such
as rubbing or grooving the walls along a particular direction
$\boldsymbol{\hat{\textbf{m}}}$, \textit{i.e.} specifying
$\ten{\Sigma}_0\sim\boldsymbol{\hat{\textbf{m}}}
\boldsymbol{\hat{\textbf{m}}}$. This would give very small $\xi$ and
effectively Dirichlet boundary conditions, as used by Cook and
co-workers \cite{Cook.Rossi04,bhavearmstrong91}.

To determine the extrapolation length an estimate for the diffusion
constant $\mathcal{D}$ is necessary. Two physical effects that can
lead to non-local dynamics and hence stress diffusion are (1) the
semi-flexibility of wormlike micelles \cite{liu93} and (2) diffusion
of micelles that carry stress \cite{bhavearmstrong91}. In the former
case the characteristic length scale is set by the persistence length
$\ell_p$.  In Appendix \ref{sec:diffusionconstant} we estimate these
two contributions to be
\begin{equation}
{\mathcal D} \tau=
\begin{cases}
\displaystyle{\frac{\ell_p^2}{126}}&\textrm{(semi-flexiblility)}\\[8truept]
D_{\textrm{tr}}\tau&\textrm{(translational diffusion)},
\end{cases}
\end{equation}
where $D_{\textrm{tr}}$ is the translational diffusion coefficient.

If we consider effects due to semiflexibility, combined with the
estimate for $W$ due steric wall interactions, we  estimate
anchoring length to be
\begin{equation}
  \label{eq:6}
  \xi\simeq\frac{\ell_p^2}{126R_g}.
\end{equation}
For typical giant micelles for which $L\simeq200\,\mu\textrm{m},
\ell_p\simeq20\,\textrm{nm}$, we find
$R_g\simeq18\ell_p\simeq670\textrm{nm}$, or
$\xi\simeq0.05\textrm{\AA}$. The interfacial width in this case
is $\ell\simeq0.09\ell_p\simeq180\textrm{\AA}$, which would imply
strong anchoring.

Alternatively, if we naively apply the estimate for the contribution
of translational diffusion to semidilute micellar solutions with
$D_{\textrm{tr}}\simeq10^{-8}\,\textrm{cm}^2/\textrm{s},
\tau\simeq1\,\textrm{s}$, one finds ${\mathcal
  D}\tau\simeq1\,\mu\textrm{m}^2$, leading to $\xi/\ell\simeq0.8\,.$
Hence in this case one might expect the effective boundary conditions
to be somewhere between Neumann and Dirichlet.

We stress that these estimates are necessarily very crude, and should
ultimately be replaced by more precise calculations.
\section{Method of Calculation}\label{sec:numerical-methods}

\subsection{Non-dimensional parameterization}
For both models we express all stresses in units of the modulus $G$,
and express time in units of the micellar (polymer) stress relaxation
time $\tau$. In cylindrical Couette flow the natural scale for length
is the quantity $R_1q$, which for small $q$ is identical to the gap
size $R_2-R_1 = R_1(1-e^{-q})$. Hence the dimensionless quantities
are:
\begin{subequations}
\begin{align}
  \hat{t} &= t / \tau&
  \hat{\dot{\gamma}} &= \dot{\gamma} \tau\\
  \hat{\ten{\Sigma}} &= \frac{\ten{\Sigma}}{G}& \hat{\mathcal D} &=
  \frac{{\mathcal D} \tau}{R_1^2 q^2}
\label{eqn:dlessdiff}\\
\hat{W}&=\frac{W}{R_1q}.
\end{align}
\end{subequations}
In cylindrical Couette flow the boundary conditions are
\begin{eqnarray}
-\hat{\mathcal D}\,\frac{\partial}{\partial x}
\ten{\Sigma}+ \hat{W}
(\ten{\Sigma}-\ten{\Sigma}_0) &=& 0 \textrm{ at } x = 0 
\\
\hat{\mathcal D}e^{-q}\,\frac{\partial}{\partial x}
\ten{\Sigma}+ \hat{W}
(\ten{\Sigma}-\ten{\Sigma}_0) &=& 0\textrm{ at } x=1,
\label{eq:mixedCouette}
\end{eqnarray}
where the change in sign arises from the orientation of the surface
normal $\mathbf{\hat{n}}$. 

We use $\epsilon\!=\!0.05, a\!=\!0.3$ for the DJS model and $\epsilon
= 0.01, \beta\!=\!0$  for the DRP model, which places the latter
in a non-monotonic regime of parameter space
(Fig.~\ref{fig:betaepsilon}).  For the shear rate startup calculations
shear rates of $\langle\hat{\dot{\gamma}}\rangle\!=\!3.8$ (DJS model)
and $\langle\hat{\dot{\gamma}}\rangle\!=\!9$ (DRP model) were used,
which are approximately in the middle of the respective stress
plateaux.
\subsection{Numerical Methods}
We solve the creeping flow equation (Eq.~\ref{eqn:gammasub}) for the
steady state banding profiles of the two models (Eqs.~\ref{eq:DJS} and
\ref{eq:DRP}) for different imposed boundary conditions and parameter
values. In all cases an average shear rate
$\langle\dot{\gamma}\rangle$ is imposed, so the local shear rate is
eliminated using Eq.~(\ref{eqn:gammasub}) to obtain a set of coupled
second order partial differential equations for $\Sigma_{rr}$,
$\Sigma_{\theta\theta}$ and $\Sigma_{r \theta}$. This set of equations
is then solved subject to the chosen boundary conditions by evolving
them using the Crank-Nicolson algorithm \cite{numC}. 

Two protocols were followed; the first for shear rate sweeps to
calculate flow curves, and the second to calculate shear rate startups
from rest. In the first protocol, shear rate sweeps, an initial
viscoelastic stress profile was generated by a set of the $20$ longest
wavelength Fourier modes that fit the desired boundary conditions.
Randomly assigned amplitudes for modes of the stress components
$\ten{\Sigma}_{\alpha\beta}$ were taken from a uniform distribution
such that the maximum total value for any component $1$. The equations
of motion were then evolved to find the steady state for the first
average imposed shear rate in the sweep, which was on either the low
or high shear rate flow branch of the constitutive curve.  This state
was then used as the initial condition for the next average shear rate
in the sweep (either incremented or decremented).  This process was
continued until the desired region of the flow curve had been
calculated. At each shear rate the system was evolved typically $500$
relaxation times $\tau$, using a time step of $0.05$; it was checked
that the steady state results did not change for smaller time steps.

In the second protocol a range of initial viscoelastic stress
configurations were subjected to a given step in the shear rate from
zero, and evolved to ensure that the same steady state was reached.
Using random initial conditions often produced multiple bands in this
protocol; these multiple bands often did not anneal into two bands,
which is probably a consequence of both the slow motion of interfaces
in one dimensional systems and the fact that multiple interface
solutions are known to be locally stable when the interface width is
much smaller than the system size \cite{grindrod}. Because we are not
interested, here, in the details of this degeneracy, we used a
spatially smooth initial condition for the viscoelastic stress tensor
which would encourage the simplest band configuration commensurate with
the given boundary conditions. A variety of different smooth initial
conditions were used to ensure that the results obtained were robust.

For smaller $\hat{\mathcal D}$ and $q$ steady state could take as long
as $10000\tau$ to attain. In the shear banding regime the positions of the
interfaces between bands were monitored to ensure that they had
stopped moving. In nearly flat geometries with weak curvatures and/or
sharp interfaces ($q<10^{-4.5}, \hat{\mathcal D}< 10^{-3}$), it was
difficult to reliably determine the steady state interfacial position;
this slow dynamics of fronts in one dimension is well known
\cite{fife77}.  Hence we did not study this range of parameter space
for this protocol.

\subsection{Notation for  (Dirichlet) Boundary Conditions}
\begin{figure}[!htb]
\begin{center}
\includegraphics[width = 0.48\textwidth]{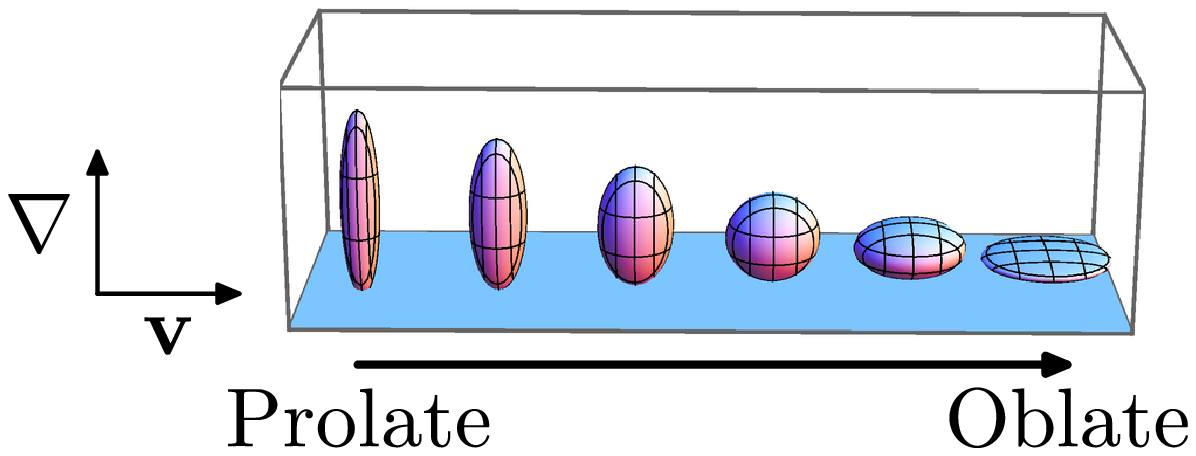}\\
\includegraphics[width = 0.48\textwidth]{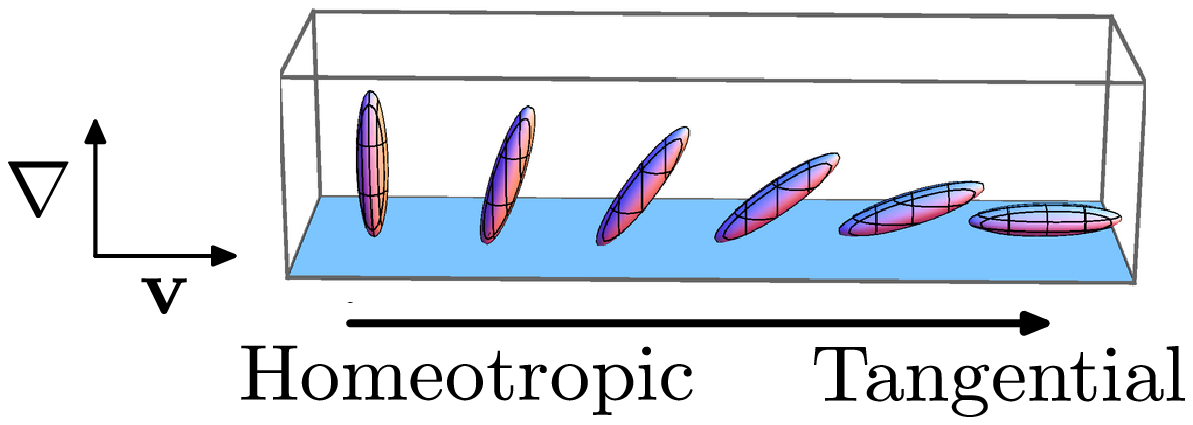}
\end{center}
\caption{Tuning between (top) prolate and oblate boundary alignment, and
  (bottom) homeotropic and tangential alignment. In (top) the principal axes
  remain unchanged and the parameters $S$ and $b$ are varied, while in
  (bottom) $S$ and $b$ are fixed to a prolate distribution, and the angle
  of orientation with respect to the surface is changed.}
\label{fig:BC}
\end{figure}

A convenient parameterization of the Dirichlet boundary condition
$\ten{\Sigma}_0$ arises from the physical interpretation of
viscoelastic stress tensor in terms of the second moment of unit
vector orientations, $\ten{\Sigma}=G\left(3 \langle
  \vec{u}\vec{u}\rangle - \ten{I}\right)$. In its principal
frame $\ten{\Sigma}_0$ can be written (in dimensionless form) as
\begin{equation}
  \tfrac{1}{3}\hat{\ten{\Sigma}_0}  +\tfrac{1}{3}\ten{I} = \left(  
\begin{array}{ccc}
      \frac{2 S+1}{3}&0&0\\
      0&\frac{1-S}{3}-b&0\\
      0&0&\frac{1-S}{3}+b
\end{array}
\right),
\label{eqn:opparams}
\end{equation}
where $S=\frac{3}{2}\langle\cos^2\theta\rangle-\frac{1}{2}$ and
$b=\langle \sin^2\theta\cos2\phi\rangle$. The conventional spherical
polar coordinates $\theta$ and $\phi$ describe the orientation of unit
vectors with respect to the principal axes.  The parameter $S$
specifies the degree of order along the principal direction, and $b$
specifies the degree of biaxiality; $S$ and $b$ obey $-\tfrac12<S<1$,
so $-\frac{1-S}{3}<b<\frac{1-S}{3}$.  The order parameter may be
illustrated by plotting an ellipsoid with principal axes parallel to
and proportional to those of the tensor (Fig.~\ref{fig:BC}).

By symmetry, we expect the principal axes of $\ten{\Sigma}_0$ to
coincide with the axes chosen, although in principle one could include
non-trivial tilt angles with respect to the surface, as is found in
the behavior of liquid crystals at some interfaces. One could also
consider, as was done in Refs.~\cite{Cook.Rossi04,rossimckinley06}, a
preferred direction of alignment along the wall. Below we will
consider  the natural principal axes set by the wall,  the
principal axes determined by the steady state bulk solutions, and a
restricted set of angles rotated with respect to the natural axes. 

\section{Review: Neumann Boundary Conditions} \label{sec:neumann}
\begin{figure}[!htb]
\begin{center}
\includegraphics[width=0.48\textwidth]{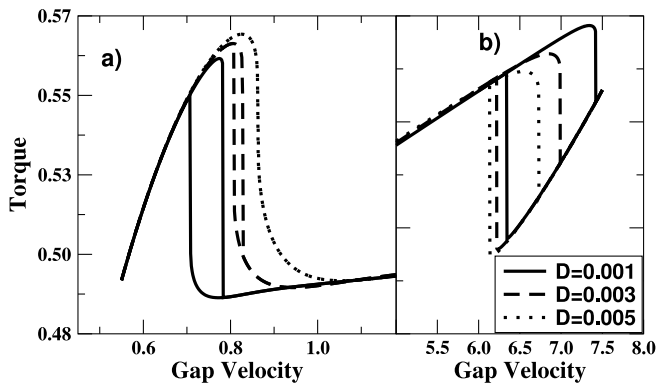}
\end{center}
\caption{Hysteresis observed in the DJS model in Couette flow for
  $q=0.00995$ upon increasing and decreasing the average shear rate
  (or the gap velocity) along the either edge of the stress
  plateau stress plateau (From \cite{olmsted99a}).}
\label{fig:hysteresisloops}
\end{figure}
We first review the results for Neumann boundary conditions,
calculated previously for the DJS model \cite{olmsted99a}. In this
case the diffusion term selects the value $T^{\ast}_{xy}$ of the total
shear stress plateau \cite{LOB00}.  At other values for the total
stress the interface moves at a speed $c\sim T_{xy}-T_{xy}^{\ast}$
\cite{radulescu99a}. In a flow geometry such as cylindrical Couette
flow, in which the total stress is inhomogeneous in steady state, the
interface lies at the position of the flow cell at which the total
stress is equal to $T_{xy}^{\ast}$ (as long as the stress gradient is
negligible over the scale of the interfacial width
$\ell=\sqrt{\hat{\mathcal D}}$) \cite{radulescu99b,radulescu99a}.
Hence the stress gradient drives the interface to the correct
position.  In a flat geometry with no stress gradient the speed of
approach to the steady state position commensurate with the imposed
average shear rate is determined by interaction with the distant wall,
and vanishes in the limit of a small diffusion coefficient
\cite{radulescu03}.

The flow curve in cylindrical Couette flow has a stress plateau with a
slope $\partial T^{\ast}/\partial\langle{\dot{\gamma}}\rangle\sim
e^q-1$, which vanishes in the planar  limit $q=0$. In either
case the flow curves display hysteresis at either edge of the stress
plateau (Fig.~\ref{fig:hysteresisloops}) \cite{olmsted99a}. During a
shear rate sweep from rest the fluid follows the low shear rate branch
to a total wall stress $T_{r\theta}>T_{xy}^{\ast}$, until a high shear
rate band forms and the stress decreases onto the stress plateau. For
a downward shear rate sweep from on the stress plateau the high shear
rate band shrinks until the interface ``touches'' the wall; at this
point a banding solution is unstable with respect to a homogeneous
phase with a higher stress on the low shear rate branch.  The width of
the hysteresis loop decreases upon decreasing the diffusion constant
$\hat{\mathcal D}$.  Hysteresis of this sort is frequently seen in
wormlike micellar solutions at the low shear rate edge of the stress
plateau \cite{GAC97,Berr97,BerrPort99}.  The nature of nucleation of a
shear banded state from a homogeneous phase is unknown and worthy of
study in its own right. Similar behavior is predicted near the high
shear rate edge of the stress plateau; this has rarely been studied
wormlike micelles, because in many cases the high shear rate band is
unstable or unattainable.
\begin{figure}[!htb]
\begin{center}
\includegraphics[width=0.45\textwidth]{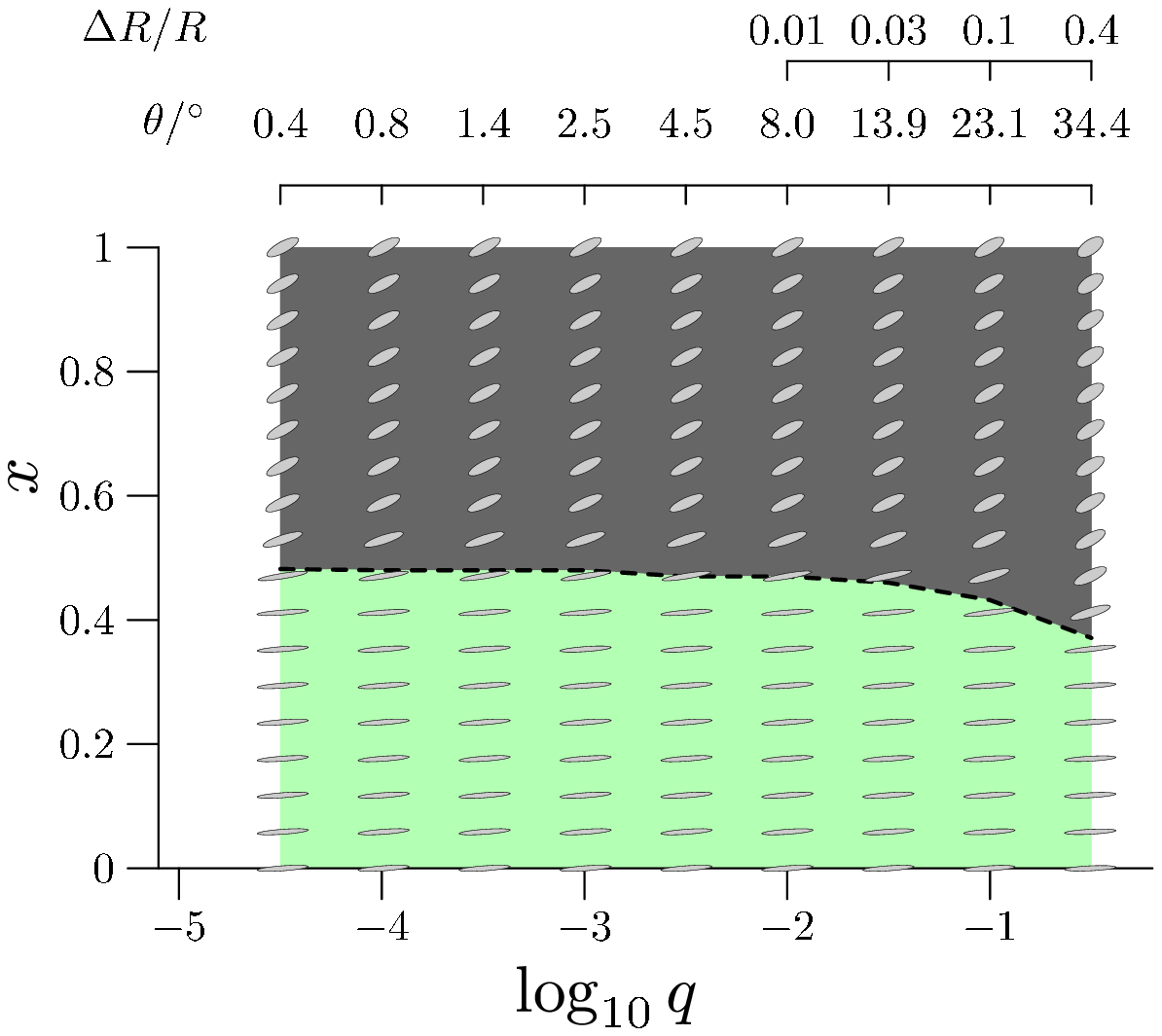}\\
\includegraphics[width=0.45\textwidth]{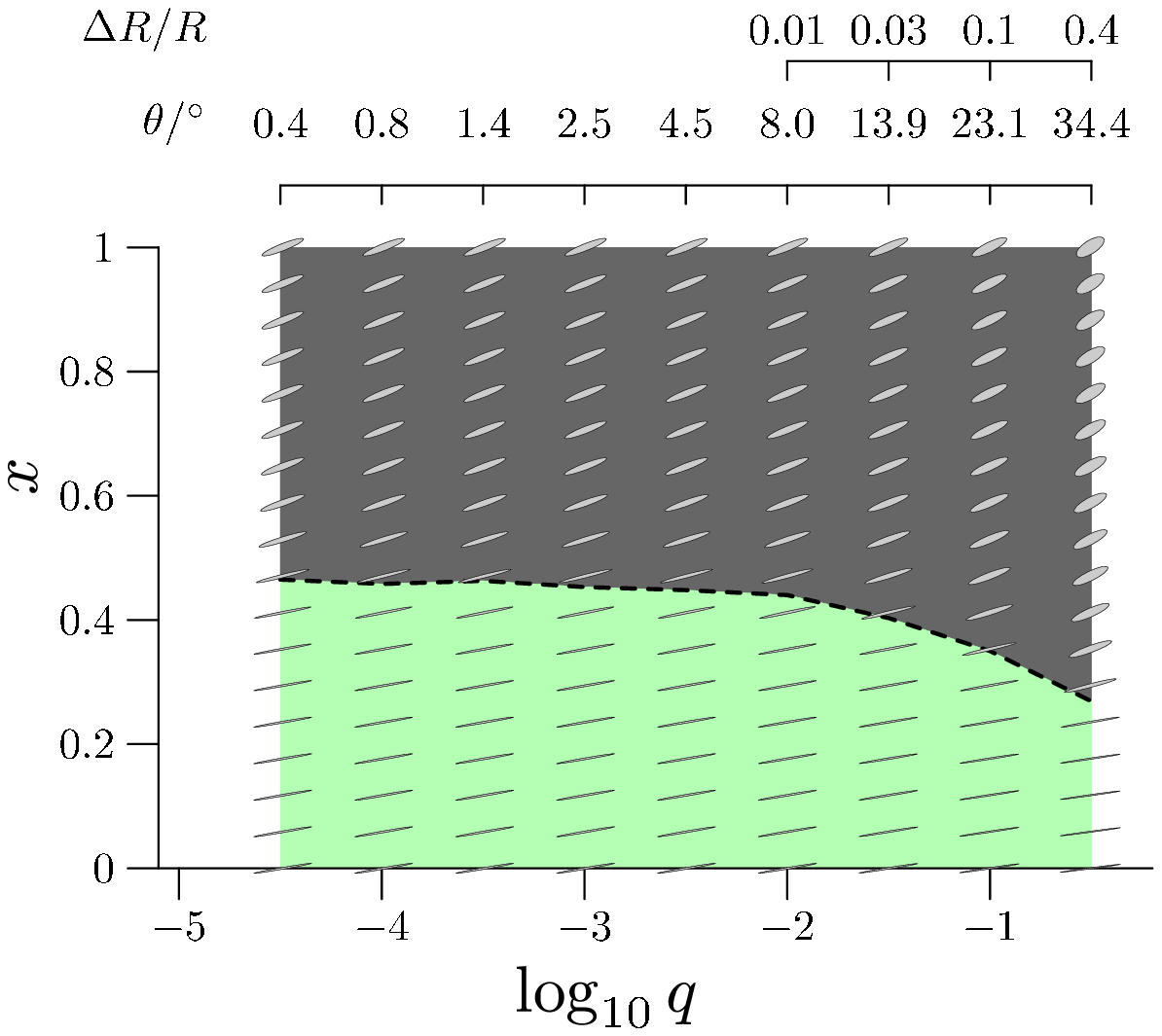}\\
 \vspace{0.3cm}
\centering\includegraphics[width=0.3\textwidth]{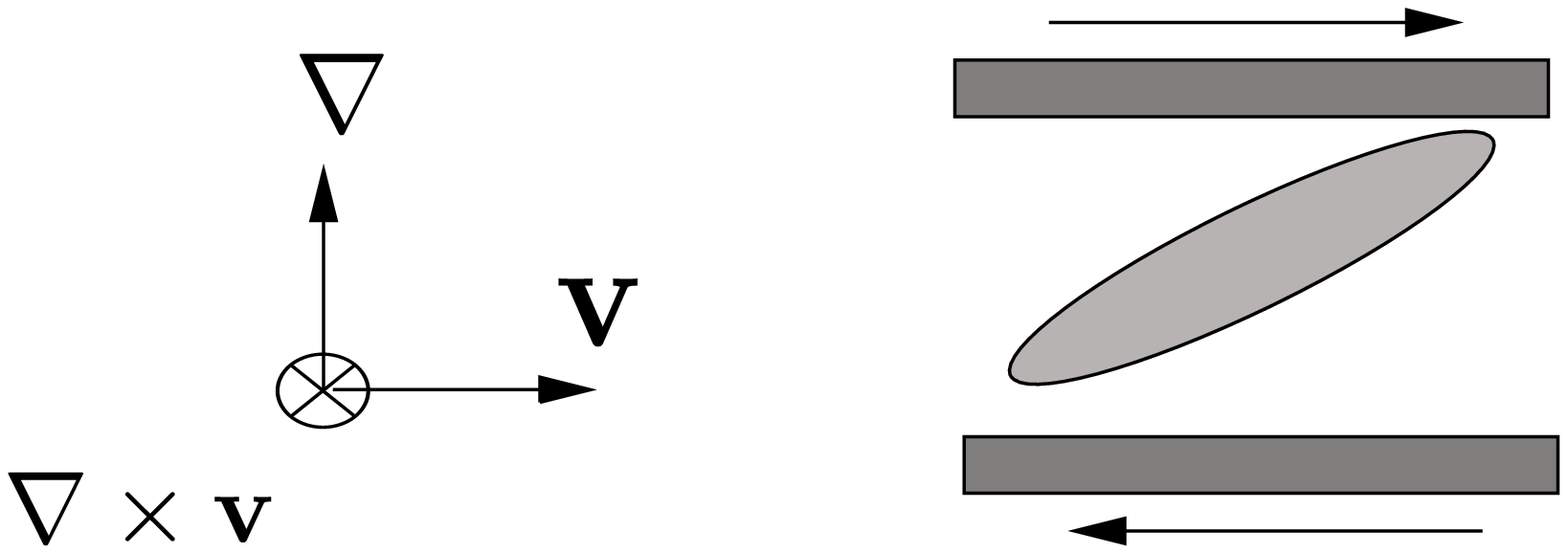}
\end{center}
\caption{Position $x$ of the interface between the shear bands for
  Neumann boundary conditions for the DJS (top) and DRP (bottom)
  models, for $\hat{\mathcal{D}} = 1.6\times10^{-3}$. The ellipses
  show the eigenvalues and principal axes of the viscoelastic stress
  tensor, which in all cases is nearly prolate in the
  velocity gradient plane. The coordinate $x$ is parallel to the
  velocity gradient, $\boldsymbol{\nabla}$.}
\label{fig:hysteresisgradient}
\end{figure}

Fig. \ref{fig:hysteresisgradient} shows the position of the
interface between bands as a function of curvature $q$, defined as the
position of midpoint between the high and low values of
$\Sigma_{r\theta}$. The high shear rate band appears to shrink at high
curvatures (large $q$) because of the total stress gradient; this
larger gradient effectively concentrates the shear rate near the inner
wall, so that the shear band occupies a slightly narrower region.  The
viscoelastic stress tensor of the high shear rate band of the DRP
model, while strongly ordered, is only slightly more aligned with the
velocity direction than in the low shear rate state. The high shear
rate state has an alignment angle of $26^\circ$ which is comparable to
that reported in experiments on wormlike micelles \cite{lerouge00}.
This contrasts with the DJS model, in which the viscoelastic stress
tensor is strongly aligned with the flow in the high shear rate band.
Hence the DRP model may be a more realistic model for wormlike
micelles than the DJS model.

The upper abscissas of Fig. \ref{fig:hysteresisgradient} shows the
stress differences in terms of geometric parameters for cylindrical
Couette flow, $\Delta R/R$ and an equivalent cone angle $\theta$ for
cone and plate flow. In Couette flow, $\Delta R/R \equiv
(R_2-R_1)/R_1=1-e^{-q}$, and the relative stress difference between
the two cylinders is $\Delta T_{r\theta}/T_{r \theta}= 1-e^{-2 q}$.
In cone and plate geometry the relative stress difference is a weak
function of the cone angle, $\Delta T_{\theta\phi}/T_{\theta
  \phi}=\tan^2 \theta$ \cite{LarsonComplex}. Hence, the cone angle
$\theta$ that is roughly equivalent to a given Couette curvature
satisfies $\tan^2\theta = 1-e^{-2q}$.

For reference, the selected stresses for the two models in a flat
geometry, in the limit of an infinite system and Neumann boundary
conditions, are:
\begin{align}
  \frac{\hat{T}^{\ast}_{r\theta}}{1+\beta}& = 
  0.62431 
&\textrm{DRP model}\\[8truept]
{\hat{T}^{\ast}_{r\theta}}\sqrt{1-a^2}& = 
  0.4829 
&\textrm{DJS model.}
\end{align}

\section{Results} \label{sec:results}

We first consider weakly curved Couette flow, and study the effect of
different Dirichlet boundary conditions and the magnitude of the
diffusion coefficient, or equivalently the widths of interfaces, on
the flow curves and attendant hysteresis. Then we study how the stress
profiles and configuration of bands depends on the competition between
the stress gradient due to the curvature of Couettte flow and the
boundary conditions. Finally, we consider mixed boundary conditions,
in which the value at the wall can adjust depending on the competition
between boundary parameters and the stress gradient.
\begin{figure*}[!htb]
\begin{center}
\includegraphics[width = \textwidth]{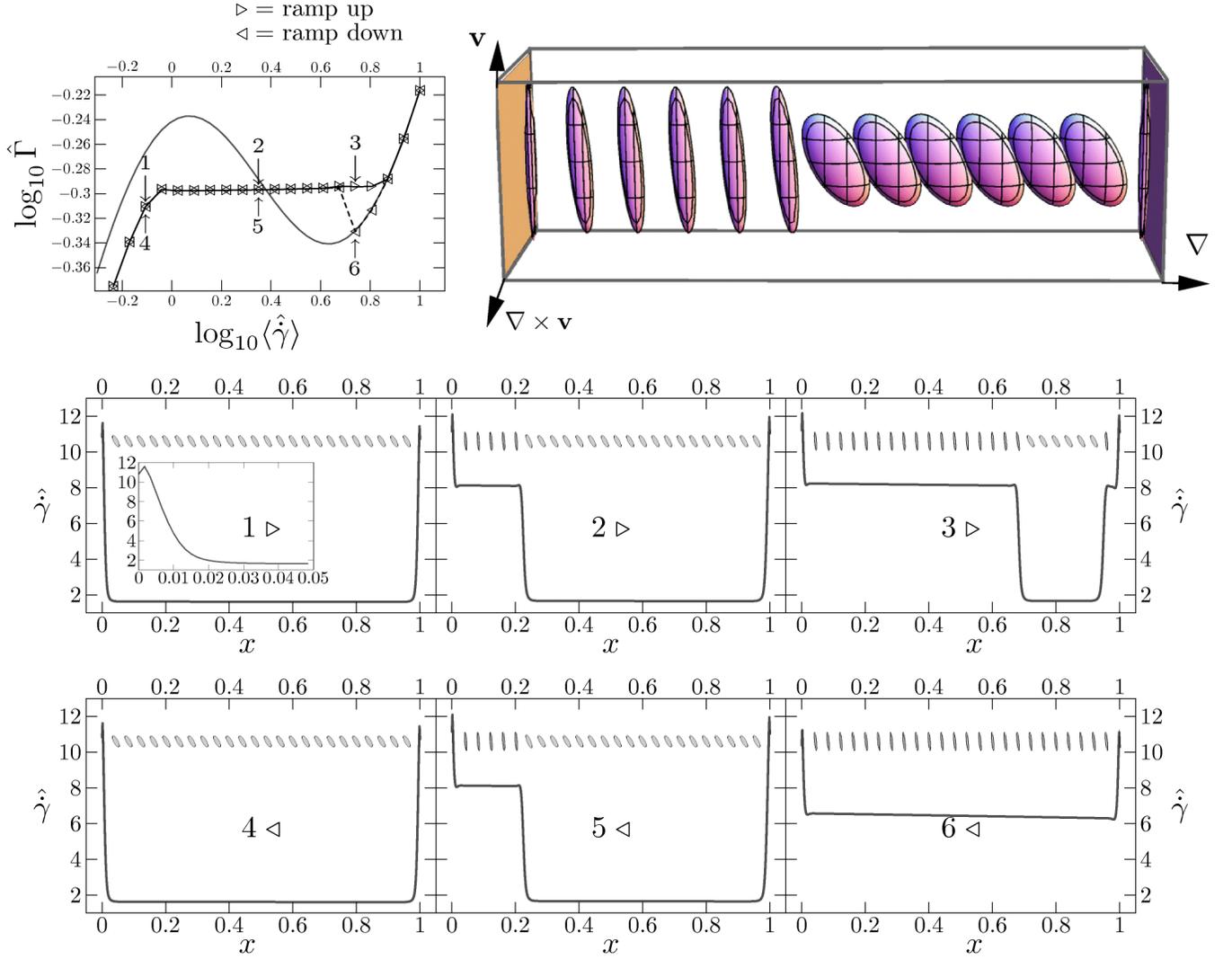}
\end{center}
\caption{Top left: constitutive curve (solid line) and flow curves
  (triangles) for upward and downward sweeps of the DJS model with
  $\hat{\mathcal{D}} = 4.4 \times 10^{-5}$, in cylindrical Couette
  flow with $q = 0.005$. The boundary condition is
  $\ten{\Sigma}_0=\textrm{diag}(2,-1,-1)$, as illustrated by the
  ellipsoids (top right). For the noted average imposed shear rates,
  figures (1)-(6) show the steady state shear rate profiles
  $\hat{\dot{\gamma}}(x)$, and cross-sections of the ellipsoids
  corresponding to $\ten{\Sigma}(x)+\Id$ in the velocity-flow gradient
  plane. In all cases the ellipsoids are nearly prolate. The inset in
  (1) shows the detail of the shear rate profile near the inner wall
  $x=0$.  }
\label{fig:HdistSBJS}
\end{figure*}
\subsection{Effect of different Dirichlet boundary conditions for
  fixed geometry } 
\subsubsection{Prolate  and flow-aligning anchoring}

Here we choose the degree of curvature to be $q=0.005$, which
corresponds to radii $R_2\simeq1.005R_1$, and enforce
$\ten{\Sigma}_0$, the value of the viscoelastic stress at the
boundaries (Dirichlet boundary conditions).  The principal axes of
$\ten{\Sigma}_0$ are chosen to be parallel to the flow, gradient and
vorticity directions.  The parametrisation of Eq.~(\ref{eqn:opparams})
was then applied so that $\ten{\Sigma}_0$ is completely aligned in the
flow direction, tangential to the walls, for $S \rightarrow1$.

Figure \ref{fig:HdistSBJS} shows the flow curves and shear rate
profiles for upward ($\triangleright$) and downward ($\triangleleft$)
sweeps of the DJS model for flow-aligning prolate boundary conditions,
$S=1.0,b=0$, similar to the value $\ten{\Sigma}_H$ of the high shear
rate branch. For low shear rates the stress is lower than that of the
low shear rate constitutive curve; profiles 1 and 4. This can be
traced to a lubricating layer (evident in the inset of profile 1)
induced by the wall, which has a higher shear rate and hence reduces
the stress for a given imposed shear rate. Hysteresis is not seen at
the low shear rate edge of the plateau, presumably because of this
lubricating layer.  Hence the effect of the boundary condition is
similar to heterogeneous nucleation or a wetting: it provides a site
upon which the shear band can easily grow, and can eliminate the
hysteresis so that the shear band grows smoothly from the wall without
requiring ``nucleation''.  Hysteresis is only seen at the high shear
rate side of the stress plateau, as shown in profiles 3 and 6, which
are at the same imposed shear rate but after different histories.
This behavior should be contrasted with Neumann boundary conditions,
in which hysteresis was predicted at both edges of the stress plateau.

In the stress plateau region the high shear rate band lies near the
inner cylinder (near $x=0$), for both increasing and decreasing shear
rate sweeps (profiles 2 and 5). Hence the higher total stress near the
wall induces preferential growth of the inner cylinder's lubricating
layer upon increasing the average shear rate, while the low shear rate
band preferentially forms at the outer wall upon decreasing the
average shear rate.  

\begin{figure*}[!htb]
\begin{center}
\includegraphics[width = \textwidth]{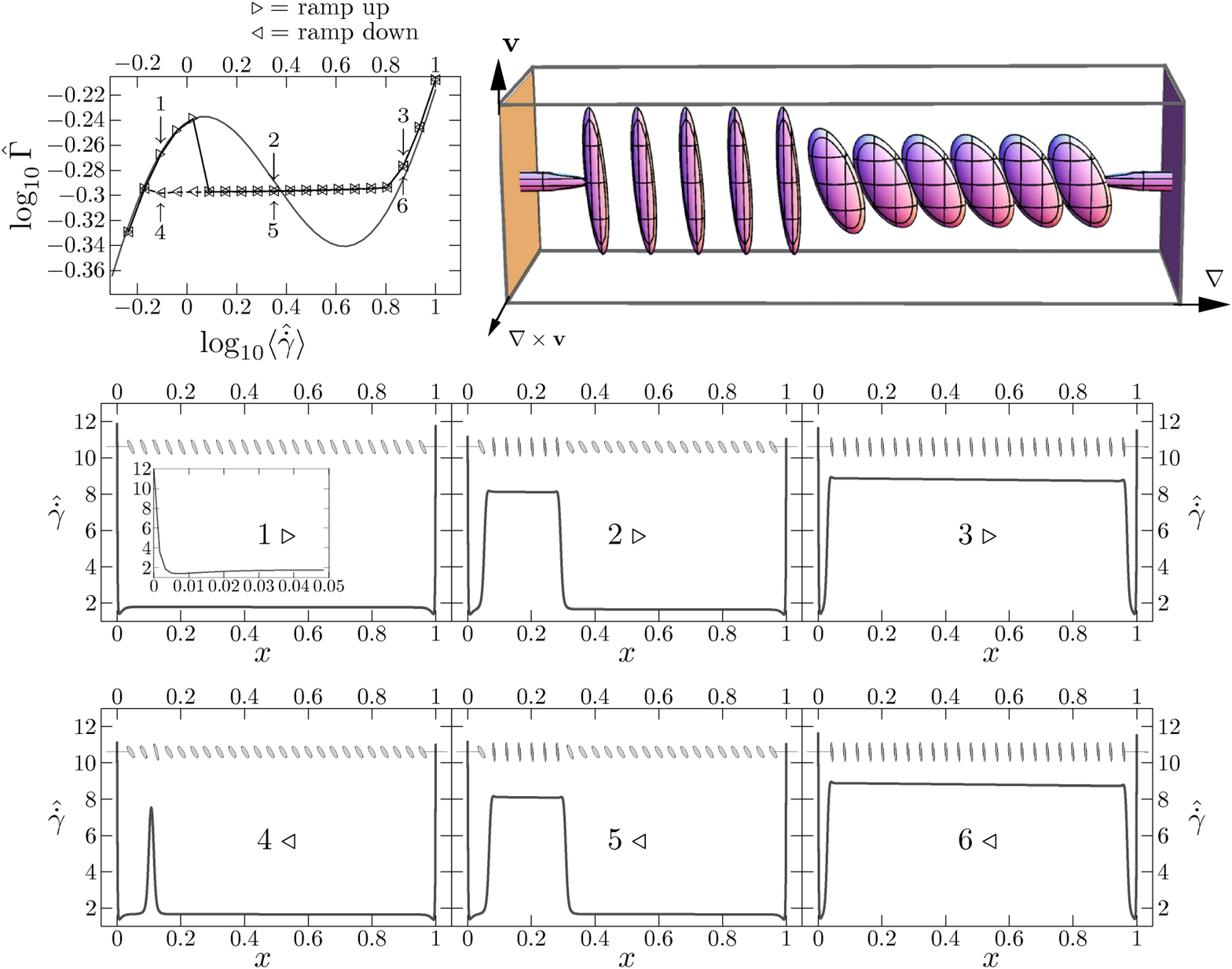}
\end{center}
\caption{Top left: constitutive curve (solid line) and flow curves
  (triangles) for upward and downward sweeps of the DJS model with
  $\hat{\mathcal{D}} = 4.4 \times 10^{-5}$, in cylindrical Couette
  flow with $q = 0.005$. The boundary condition is
  $\ten{\Sigma}_0=\textrm{diag}(-1, 2, -1)$, as illustrated by the
  ellipsoid (top right). (1)-(6) show the shear rate profiles
  $\hat{\dot{\gamma}}$, and cross-sections of the ellipsoids
  corresponding to $\ten{\Sigma}(x)+\Id$ in the velocity-flow gradient
  plane. Except for very near the boundaries, the ellipsoids are
  nearly prolate. The inset in (1) shows the detail of the shear rate
  profile near the inner wall $x=0$. }
\label{fig:LdistSBJS}
\end{figure*}
\subsubsection{Homeotropic anchoring}
The converse behavior is observed when the boundary conditions are
less favorable to the high shear rate phase. Fig.~\ref{fig:LdistSBJS}
shows the flow profiles and flow curves for $S=-0.5, b=-0.5$, which
corresponds to homeotropic orientation with principal axis in the
gradient direction. In this case a hysteresis loop is found at the low
shear rate end of the plateau (profiles 1 and 4), and not at the high
shear rate end (profiles 3 and 6). This boundary condition induces a
more viscous layer of low shear rate material near the wall, even well
into the stress plateau (profiles 2 and 5). Profiles 2 and 5 are not
quite superposable, indicating that steady state was probably not
exactly attained for these profiles; indeed, this value $\hat{\mathcal
  D}=4.4\times 10^{-4}$ is at the lower limit of our computational
capabilities. The more viscous layer near the boundary increases the
total stress on the high shear rate branch above than that of the
constitutive curve (profiles 3 and 6), as opposed to the lubricating
layer found for the flow-aligning boundary condition.
\begin{figure*}[!htb]
\begin{center}
\begin{tabular}{cc}
  \subfigure[Parameter space and characteristic flow curves:
  DJS  Model.] {
\includegraphics[width = 0.477\textwidth]{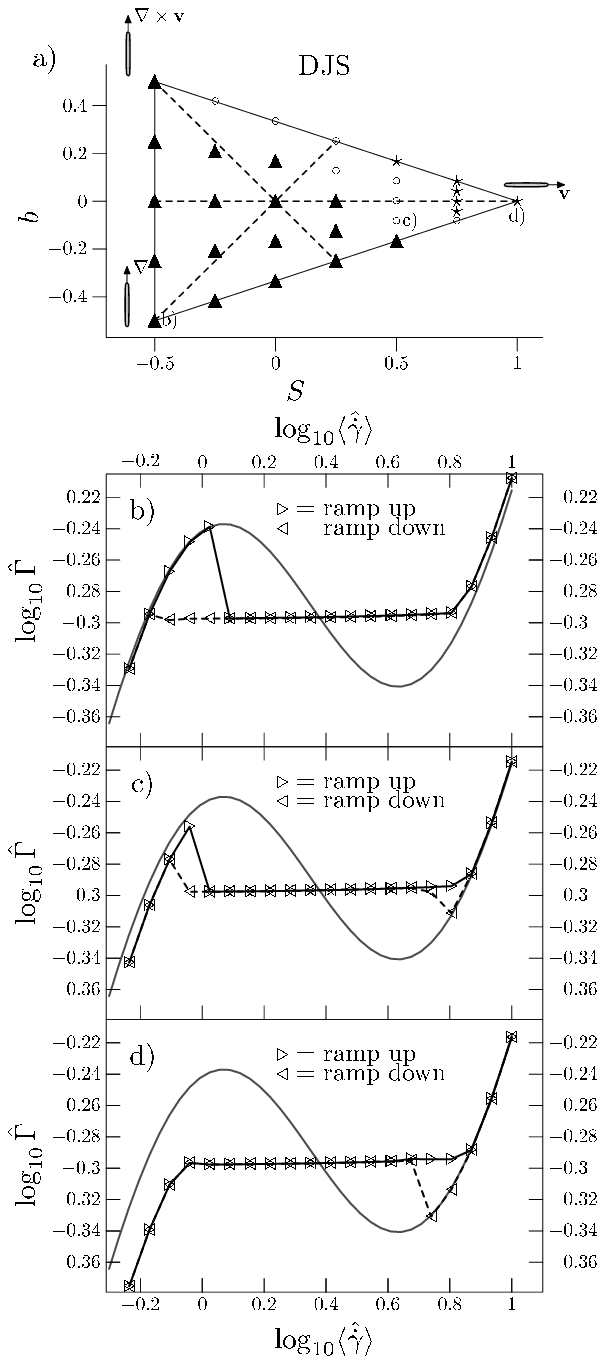}
\label{fig:mapJS}
}
&
\subfigure[Parameter space and characteristic  flow curves: DRP
Model.]{  
\includegraphics[width = 0.5\textwidth]{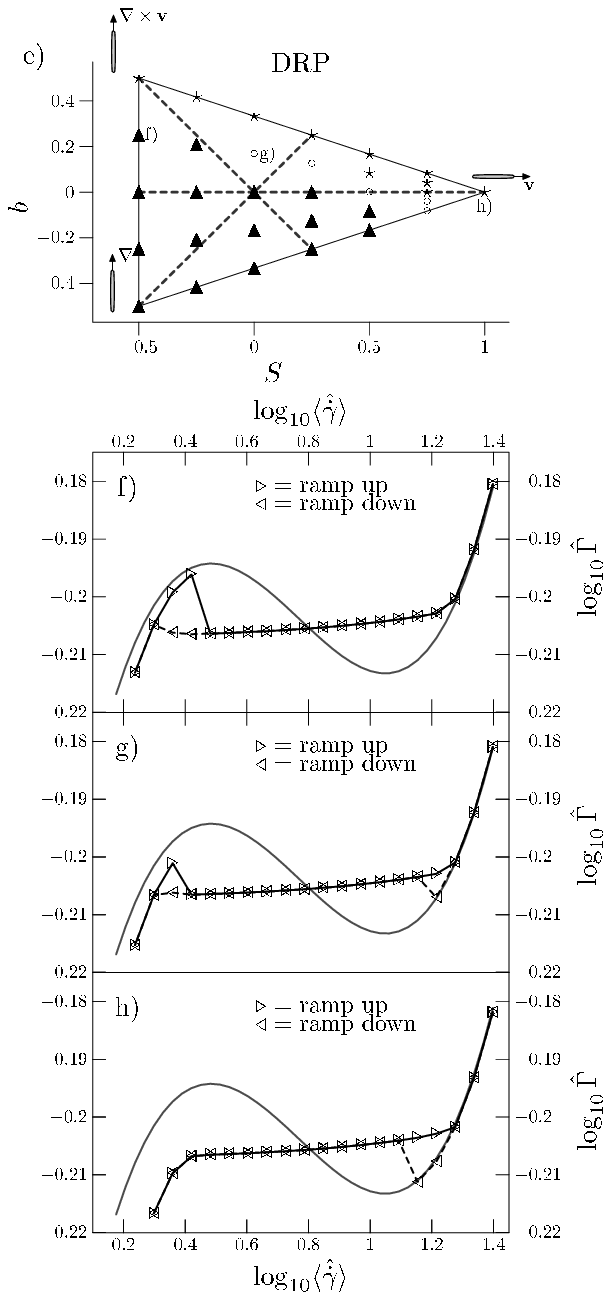}
\label{fig:mapRP}
}
\\
\\[1ex]
\end{tabular}
\end{center}

\vskip-1.0truecm
\caption{(a,e) Regions of $S-b$ parameter space for $\ten{\Sigma}_0$
  that give rise to different signatures of hysteresis during shear
  rate ramps, for $q = 0.005$, and $\hat{\mathcal{D}} = 4.4 \times
  10^{-5}$. The dashed lines on (a,e) correspond to uniaxial order
  interpolating between an oblate ``pancake'' and a prolate ``needle''
  (at $S=-0.5, b=\pm 0.5$, and $S = 1$), aligned along the vorticity,
  flow gradient, and flow directions. Hysteresis occurs at the low
  shear rate end of the stress plateau ( $\blacktriangle$, b,f); at
  the high shear rate end of the stress plateau ($\star$, d,h); or at
  both ends shear rate end of the stress plateau ($\circ$, c,g).}
\label{fig:FCSBJS}
\end{figure*}

The shear rate adjacent to the wall is extremely high (inset to
profile 1), even higher than that of the high shear rate phase. This
is because the principal axes of the boundary condition
$\ten{\Sigma}_0$ are aligned with the velocity, gradient, and
vorticity, such that the shear component $\ten{\Sigma}_{0r\theta}$
vanishes. To compensate for this vanishing contribution to the shear
stress, a very high shear rate is required to obtain the necessary
value for the total shear stress. For boundary conditions conducive to
the low shear rate phase, the viscoelastic stress tensor
$\ten{\Sigma}$ adjacent to the wall nonetheless ``heals'' to the value
characteristic of the low shear rate branch, whereas in the
flow-aligning case it heals to the value characteristic of the high
shear rate branch.  Although our numerics can adequately resolve the
sharp gradient near the wall, it is likely that higher order gradients
are necessary for a constitutive model to give physically accurate
results.

\subsubsection{Variation with anchoring angle and from prolate to
  oblate anchoring}
A summary of the behavior of the DJS and DRP models for a range of
boundary conditions is shown in Fig.~\ref{fig:FCSBJS}.  There are
essentially three types of behaviors: hysteresis loops can be observed
on (1) the high shear rate end of the stress plateau, (2) the low
shear rate end of the stress plateau, or (3) both ends of the stress
plateau. Hysteresis vanishes on those flow branches whose viscoelastic
stress, loosely, is similar to the imposed boundary value
$\ten{\Sigma}_0$. The two models have broadly similar qualitative
behavior. [The apparently greater slope across the plateau for the DRP
model (Fig.~\ref{fig:mapRP}) is due to the curvature of Couette flow
and the smaller vertical scale.]

The behavior was similar when the boundary conditions were altered by
choosing a prolate $\ten{\Sigma}_0$ and rotating its principal axis from
the flow gradient (homeotropic) to the velocity (tangential)
directions. For tangential boundary conditions (resembling the high
shear rate state) there was no overshoot at the start of the plateau
either on ramping the shear rate up or down. However on ramping down
the shear rate the flow curve undershoots the stress plateau at the
high shear rate end of the plateau. As the boundary condition is
rotated the hysteresis at the high shear rate end vanishes, and starts
to develop at the low shear rate end. For homeotropic alignment
(resembling the low shear rate state) the hysteresis completely
vanishes at the high shear rate end of the plateau, and hysteresis
develops at the low shear rate end.

\begin{figure*}[!htb]
\begin{center}
\includegraphics[width=0.9\textwidth]{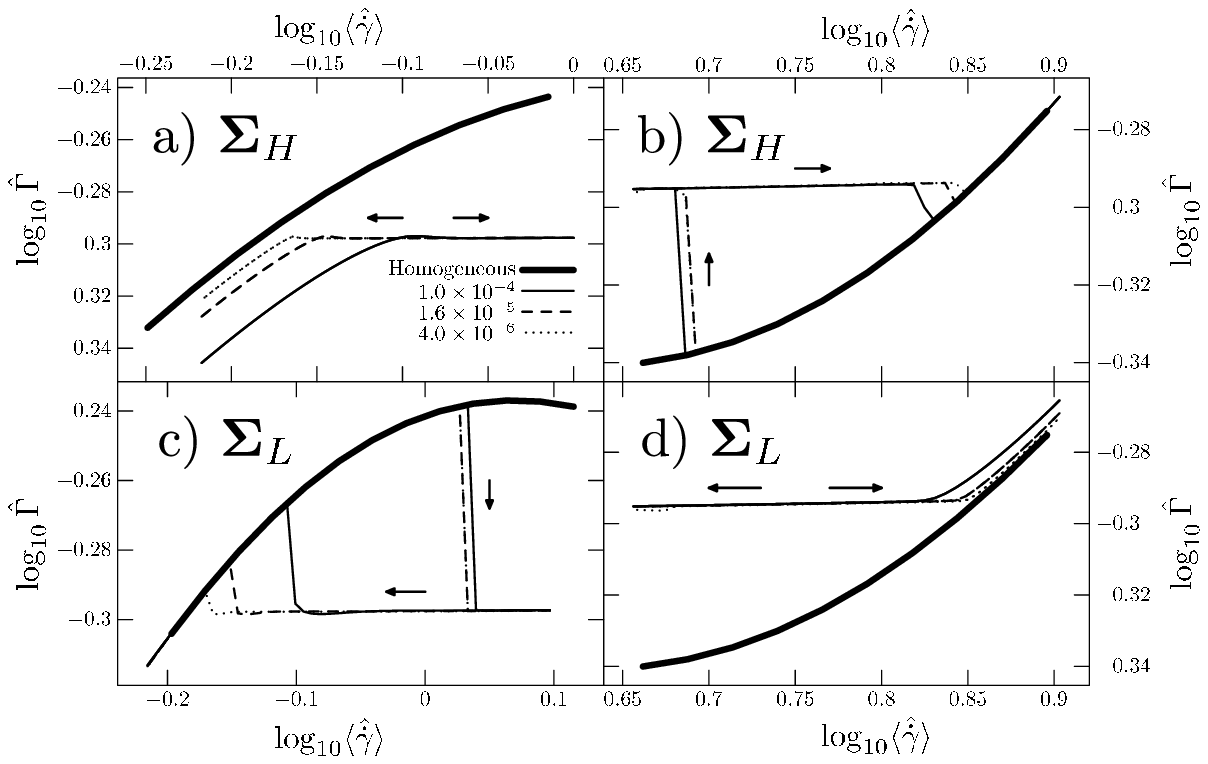}
\end{center}
\caption{Constitutive curve and flow curves for the DJS model, for
  $q=0.005$, and diffusion constants $\hat{\mathcal{D}} = 1.0
  \times10^{-4}$ (solid), $1.6 \times10^{-5}$ (dashed) and $4.0 \times
  10^{-6}$ (dotted).  (a,c) show the start of the plateau, while (b,d)
  show the end; and the boundary condition $\ten{\Sigma}_0$ equal to
  the viscoelastic stress $\ten{\Sigma}_H$ or $\ten{\Sigma}_L$ of,
  respectively, the high (a,b) or low (c,d) shear rate branch. The
  arrows indicate the direction in which the hysteresis loops are
  encircled (b,c) or the shear rate swept (a,d).}
\label{fig:FCHLJS}
\end{figure*}

\subsubsection{Effect of the diffusion constant}
Having determined that similarity of the imposed viscoelastic stress
$\ten{\Sigma}_0$ to that of either the high or low shear rate phases
is crucial, we focus for the remainder of this paper on boundary
conditions for which $\ten{\Sigma}_0$ is one of these values. 
For the DJS model with $\epsilon=0.05$ and $a=0.3$ these values are
\begin{widetext}
  \begin{subequations}
    \begin{align}
      \ten{\Sigma}_L:&& (\Sigma_{L\theta\theta},
      \Sigma_{Lrr},\Sigma_{Lr\theta})&=(0.40,-0.21,0.47)&&\textrm{low
        shear rate branch}\\
      \ten{\Sigma}_H:&& (\Sigma_{H\theta\theta},
      \Sigma_{Hrr},\Sigma_{Hr\theta})&=(1.4,-0.75,0.12)&&\textrm{high
        shear rate branch,}
    \end{align}
  \end{subequations}
  and for the DRP model with $\beta = 0$ and $\epsilon = 0.01$ they
  are
  \begin{subequations}
    \begin{align}
      \ten{\Sigma}_L:&& (\Sigma_{L\theta\theta},
      \Sigma_{Lrr},\Sigma_{Lr\theta})&=(0.81,-0.40,0.60)&&\textrm{low
        shear rate branch}\\
      \ten{\Sigma}_H:&& (\Sigma_{H\theta\theta},
      \Sigma_{Hrr},\Sigma_{Hr\theta})&=(1.7,-0.85,0.44)&&\textrm{high
        shear rate branch.}
    \end{align}
  \end{subequations}
\end{widetext}

The extent of the hysteresis depends on the magnitude of the diffusion
constant.  Fig.~\ref{fig:FCHLJS} shows how ${\mathcal D}$, or
equivalently the width $\ell\sim\sqrt{{\mathcal D}}$ of the interface,
influences the flow curves, for the two different boundary conditions
of the DJS model. As the diffusion constant is reduced the
heterogeneous flow curve approaches the homogeneous constitutive
curve, and the degree of hysteresis decreases. This occurs for both
the DRP (not shown) and DJS models. As noted above, this is because
the interfacial layer occupies a progressively smaller fraction of the
sample: the separation between the heterogeneous flow curve and
homogeneous constitutive curve is proportional to
$\sqrt{\mathcal{D}}\sim\ell$.  This behavior should be contrasted with
the Neumann case (Fig.~\ref{fig:hysteresisloops}), in which the flow
curve matches the constitutive curve for all ${\mathcal D}$.

\subsection{The effects of flow geometry and total stress gradient} 
\subsubsection{Dirichlet boundary conditions}
For a flow cell with a stress gradient, such as a cylindrical Couette
cell, the interface in simple banding flow with Neumann boundary
conditions lies at the position in the flow cell where the stress is
equal to the selected total shear stress $T^{\ast}_{r\theta}$.  With
Dirichlet boundary conditions there are typically three bands and two
interfaces, because of the boundary layer imposed by the wall.  We now
study how the position of these bands varies as a function of the flow
geometry and diffusion coefficient.  Calculations were performed using
the second protocol outlined in Sec.~\ref{sec:numerical-methods}:
starting up from rest with a suitable initial condition and evolving
at fixed average strain rate to steady state. The boundary condition
$\ten{\Sigma}_0$ was chosen to be either $\ten{\Sigma}_L$ or
$\ten{\Sigma}_H$.
\begin{figure*}[!htb]
  \begin{tabular}{cc}
    \includegraphics[width = 0.48\textwidth]{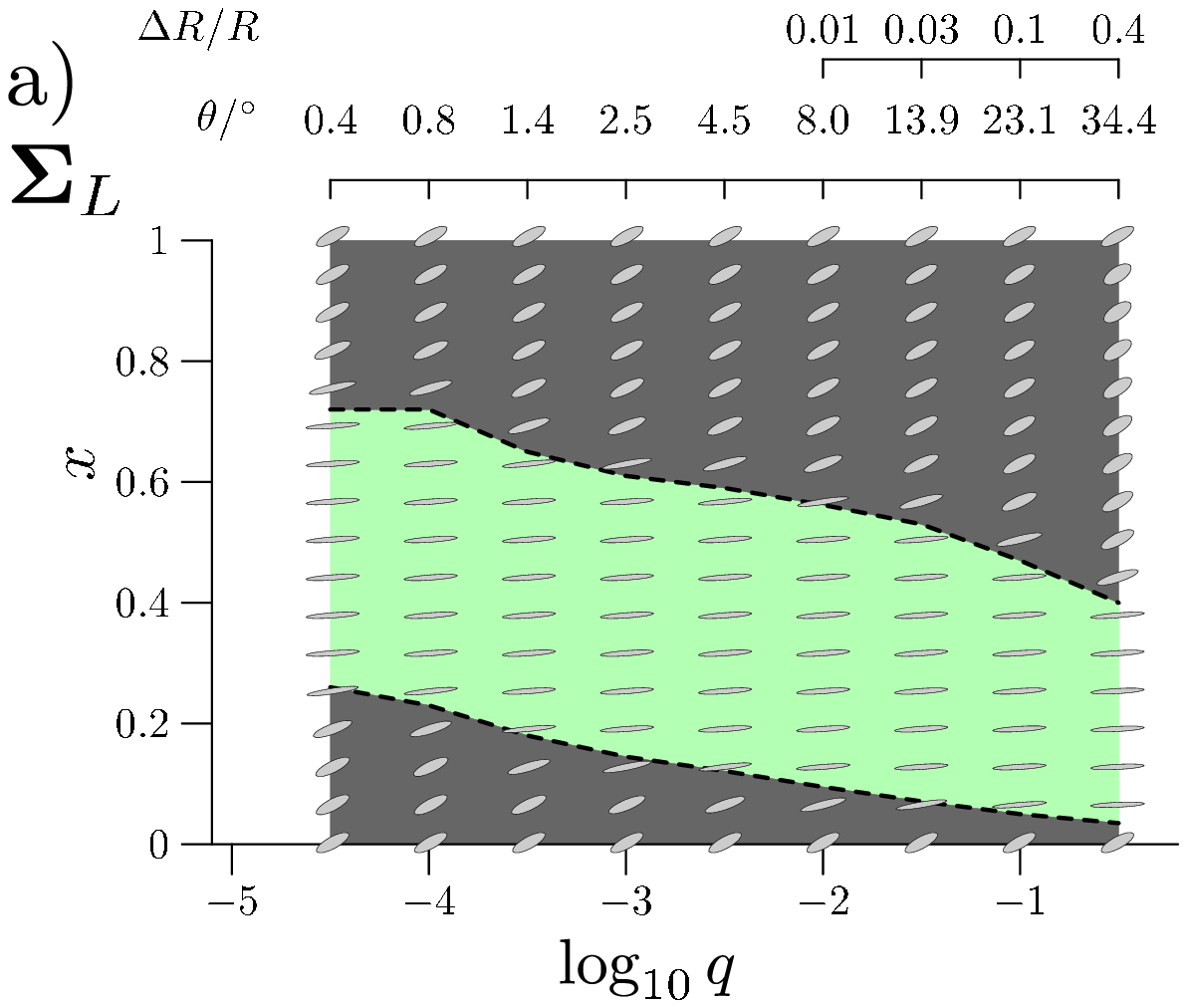}
    &
    \includegraphics[width = 0.48\textwidth]{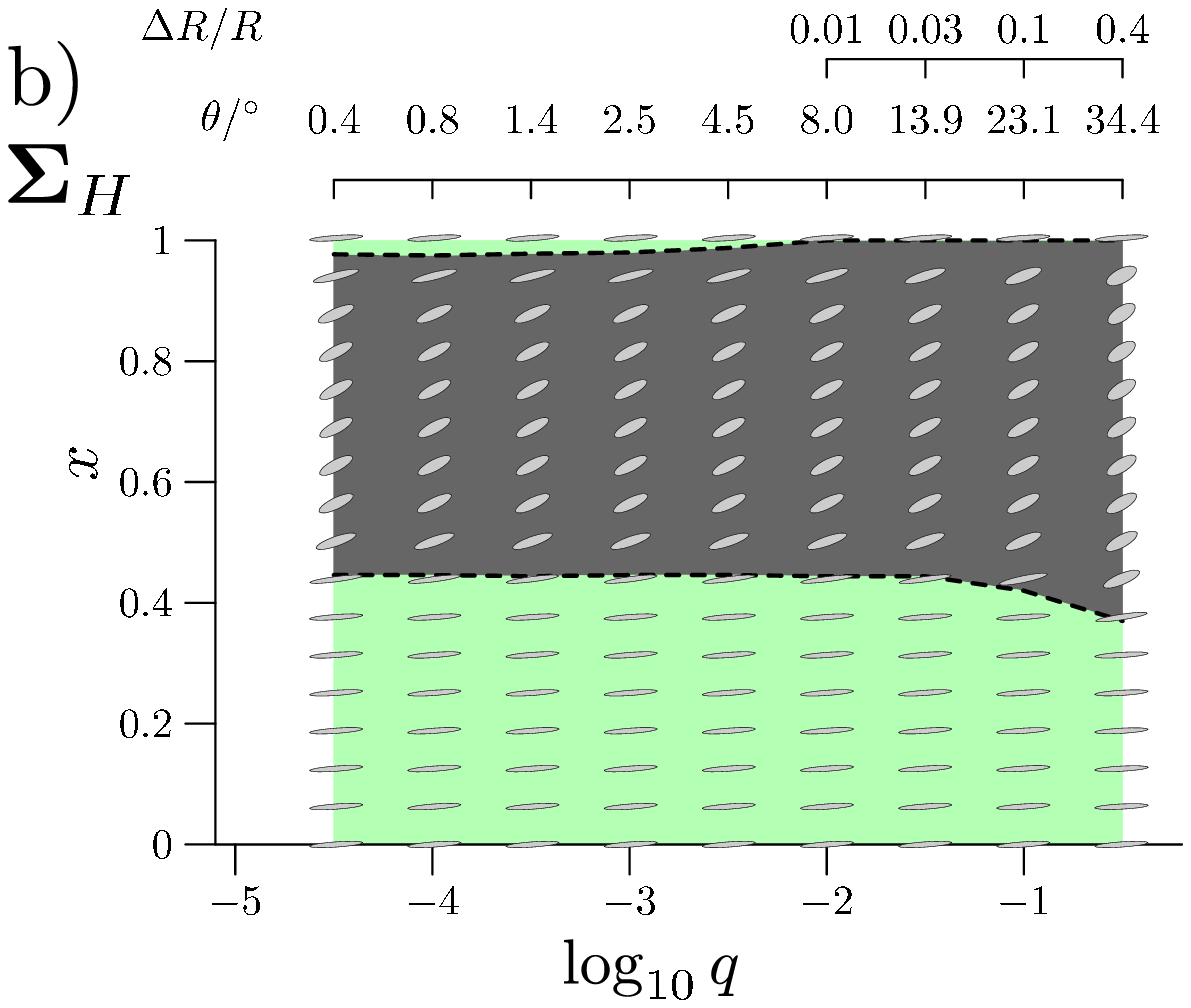}
\\
    \includegraphics[width = 0.48\textwidth]{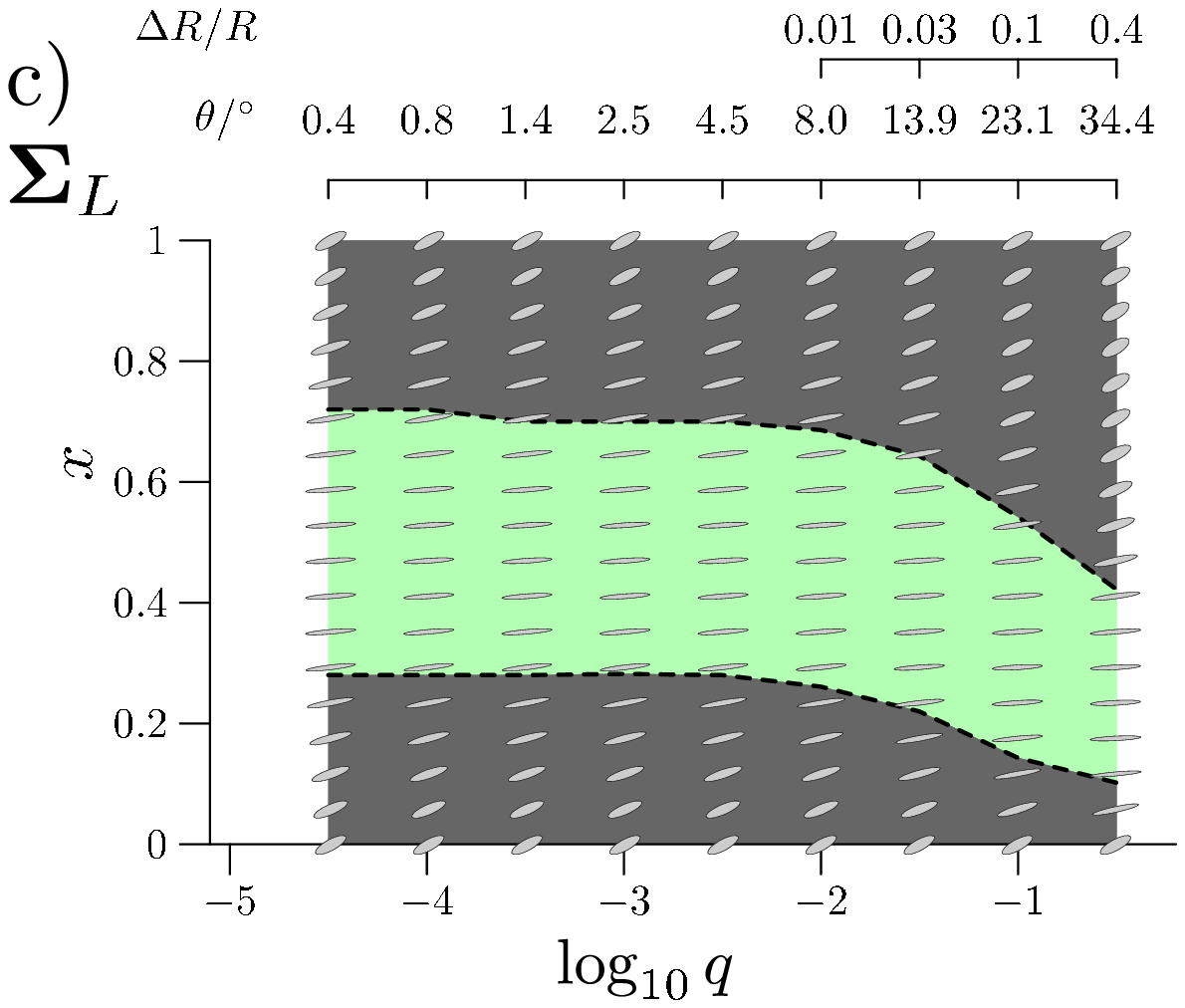}
    &
    \includegraphics[width = 0.48\textwidth]{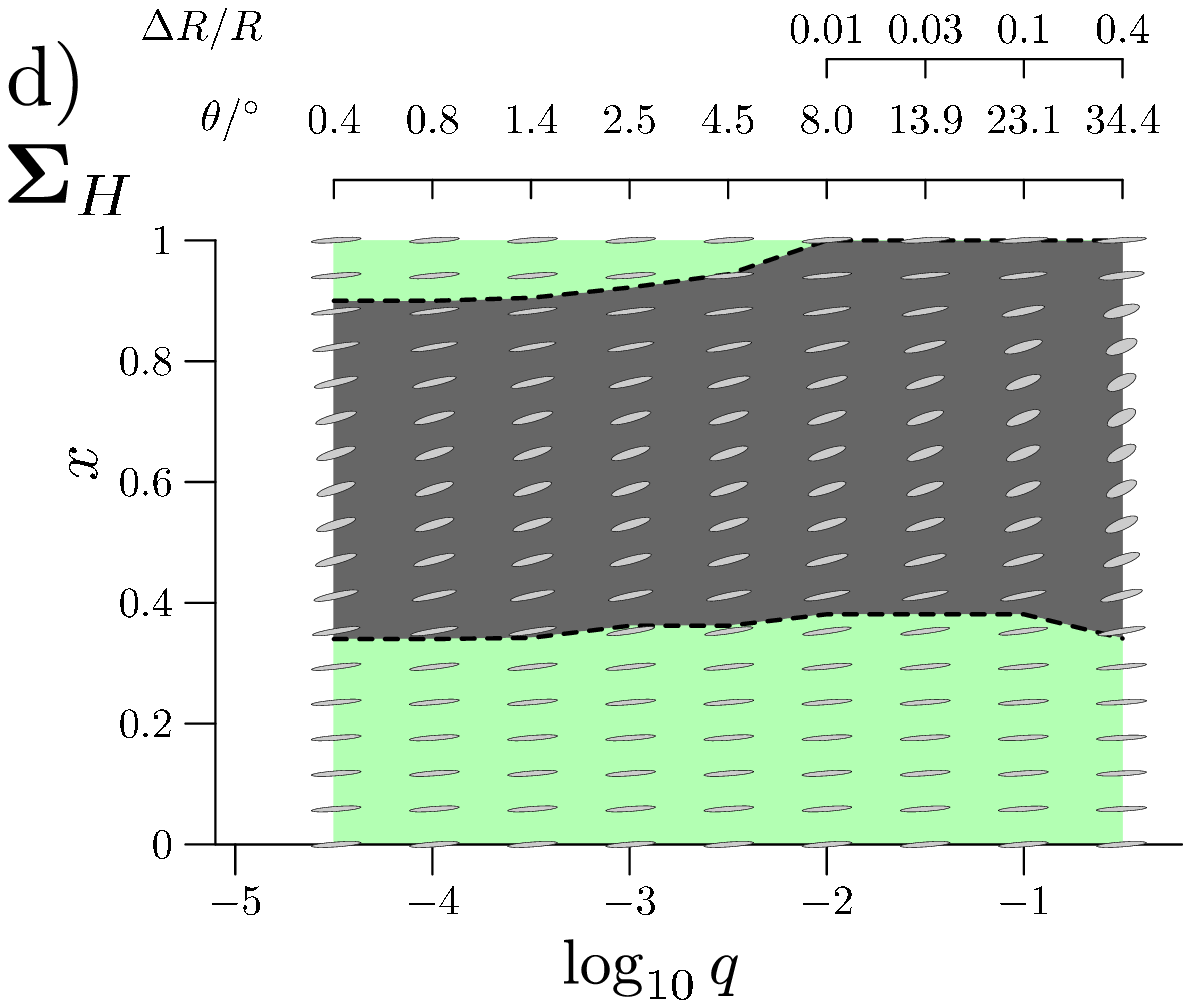}
  \end{tabular}
\caption{Viscoelastic stress ellipses as a function of position $x$
  and curvature $q$, for the DJS model with
  $\langle\hat{\dot{\gamma}}\rangle = 3.8
  (\log_{10}\langle\hat{\dot{\gamma}}\rangle=0.58)$.  The diffusion
  constant is $\hat{\mathcal{D}} = 10^{-3}$ (a,b) and
  $\hat{\mathcal{D}} = 10^{-2}$ (c,d). The high shear rate phase is
  lightly shaded (green), and the low shear rate phase is dark grey.
  The boundary conditions are $\ten{\Sigma}_0=\ten{\Sigma}_H$ or
  $\ten{\Sigma}_L$. The top axes show the equivalent Couette cylinder
  gap $\Delta R$ relative to the inner cylinder radius $R$, and
  equivalent cone angles $\theta$ if we assume that the stress
  gradient mimics that found in cone and plate flow.}
\label{fig:JSvq}
\end{figure*}

\begin{figure}[!htb]
\begin{center} 
    \includegraphics[width = 0.48\textwidth]{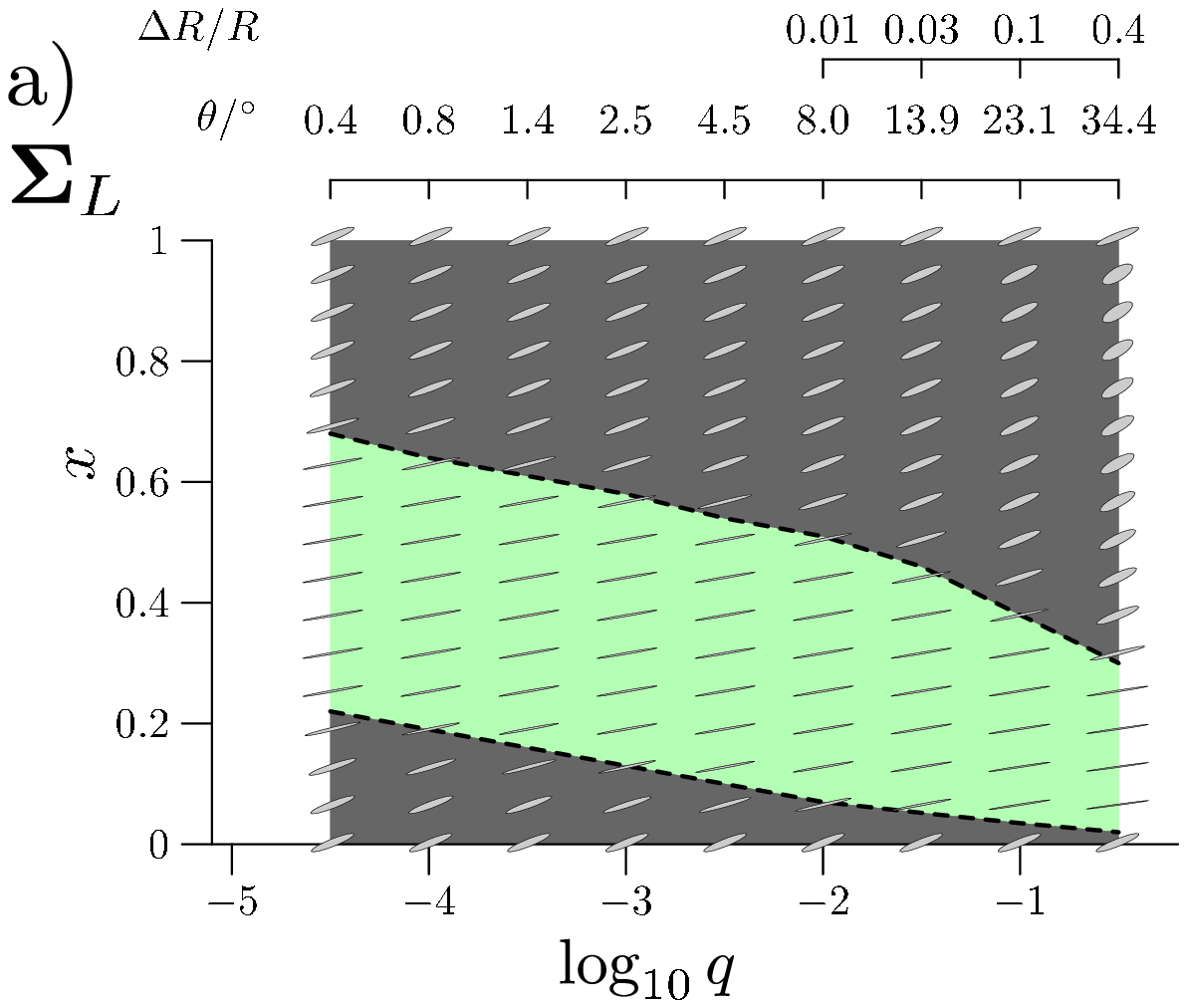}\\
    \includegraphics[width = 0.48\textwidth]{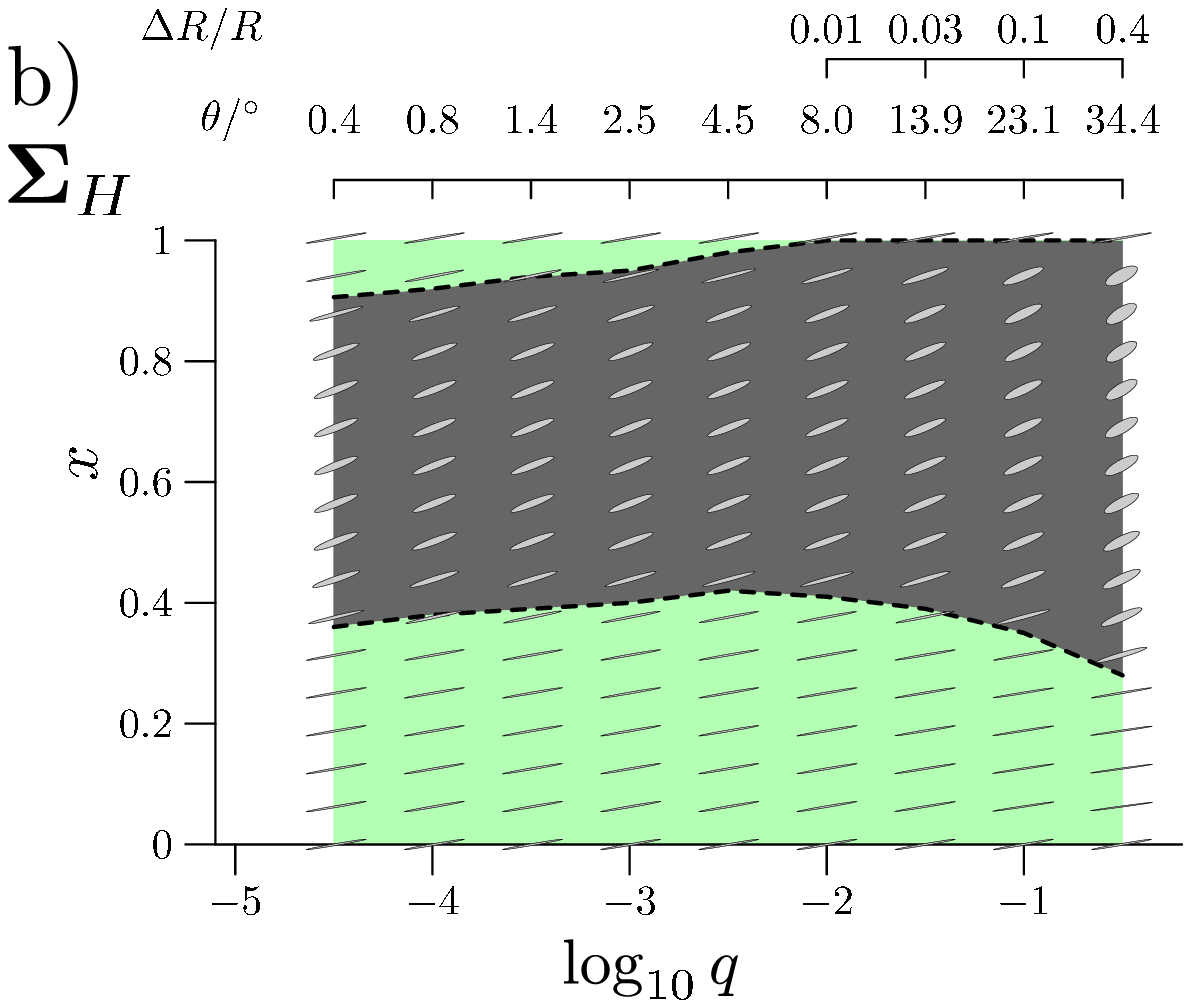}
\end{center}
\caption{Viscoelastic stress profiles, shown as ellipses as a function
  of position $x$, for the DRP model in different degrees of curvature
  $q$, for $\hat{\mathcal{D}} = 10^{-3}$ and
  $\langle\hat{\dot{\gamma}}\rangle = 9\,\,
  (\log_{10}\langle\hat{\dot{\gamma}}\rangle=0.95)$.  The high shear
  rate phase is lightly shaded (green), and the low shear rate phase
  is dark grey.  The boundary conditions are
  $\ten{\Sigma}_0=\ten{\Sigma}_H$ (right) or
  $\ten{\Sigma}_0=\ten{\Sigma}_L$ (left). The upper abscissa shows the
  equivalent geometrical parameters for cylindrical Couette ($\Delta
  R/R$) and cone and plate ($\theta$) geometries.}
\label{fig:RPvq}
\end{figure}

In a flat geometry the phase near the wall is that most similar to the
imposed boundary conditions, ideally resulting in a symmetric three
band configuration stabilized by the very weak interactions of the
interfaces with the walls.  For strong enough curvature we expect the
high shear rate phase to preferentially occupy the high stress
regions. This is indeed what we find.  For increasing curvature the
three band configuration becomes increasingly asymmetric, as seen in
both the DJS and DRP models (Figs.~\ref{fig:JSvq} and \ref{fig:RPvq}).
A crossover from three banded to nearly two-banded behavior occurs at
smaller $q$ values (weaker curvature) for smaller $\mathcal{D}$. The
three band state is most pronounced when the boundaries induce the low
shear rate phase $\ten{\Sigma}_0$.

The behavior described above can be rationalized as follows. For a
symmetric flat system ($q=0$) both interfaces lie at the selected
stress, since the total stress is uniform. The very weak effect a
slightly curved system $q\gtrsim0$ breaks the symmetry of the shear
banded structure, and only one (or neither) interface can lie at the
selected stress; at the same time, the stress gradient would favor
locating the high shear band near the inner cylinder.  These two
tendencies adjust the band configuration such that high (low) shear
rate bands grow (shrink) near the inner cylinder and shrink (grow)
near the outer cylinder. For example, for boundary conditions that
favor the low shear rate band the band near the inner cylinder should
become narrower with increasing curvature (as seen in
Fig.~\ref{fig:JSvq}a,c).  However, a larger diffusion coefficient
broadens the interfaces between bands, which means that the walls can
strongly influence the bulk behavior.  Correspondingly, for large
${\mathcal D}$ the position of the center shear band moves more
smoothly as a function of the geometric curvature $q$. For example,
compare Fig.~\ref{fig:JSvq}c ($\hat{\mathcal D}=10^{-2}$) with
Fig.~\ref{fig:JSvq}a ($\hat{\mathcal D}=10^{-3}$).

Just as for Fig.~\ref{fig:hysteresisgradient}, the effective cone
angles are again shown in the upper abscissas in Figures
\ref{fig:JSvq} and \ref{fig:RPvq}. For small diffusion constant,
Fig.~\ref{fig:JSvq}a shows that in cone and plate flow with a typical
angle of $\theta=4^{\circ}$, a three banded structure could be
observed, while the third band would be much smaller in Couette flow
with $\Delta R/R=0.05$, and could easily be interpreted as a two band
structure.  Thus, a three band structure, as observed by Britton and
Callaghan \cite{BritCall97c}, could be consistent with boundary
conditions that favor the low shear rate form of the viscoelastic
stress.

Finally, we note that these results are consistent with those of Cook
and co-workers \cite{Cook.Rossi04,rossimckinley06}, who studied a
two-fluid model of wormlike micelles.  In cylindrical Couette flow
with $q = 0.064, \hat{\mathcal D}=10^{-3}$ they also found a
three-band state, but did not explore the effects of geometry or
diffusion coefficient.

\subsubsection{Competing boundary conditions}
Finally we allow surface anchoring to compete with the bulk spatial
gradients, according to Equation (\ref{eq:mixed}). By varying $W$, or
equivalently the extrapolation length $\xi=D\tau/W$, the boundary
condition at the wall can be tuned smoothly from a fixed value
$\ten{\Sigma}_0$ as $W\rightarrow\infty\, (\xi\rightarrow0)$ to zero
gradient as $W\rightarrow0\, (\xi\rightarrow\infty)$. As above, we
consider boundary conditions in which $\ten{\Sigma}_0$ is the
viscoelastic stress of either the low or high shear rate branch,
$\ten{\Sigma}_L$ or $\ten{\Sigma}_H$. We vary both the extrapolation
length $\xi$ and the geometric curvature $q$, and monitor the behavior
of a well-formed shear band, with an imposed average shear rate such
that the high and low shear rate bands are roughly the same size. We
find qualitatively similar results for both the DJS and DRP models.

\begin{figure*}[!htb] 
  \begin{center}
\includegraphics[width = \textwidth] {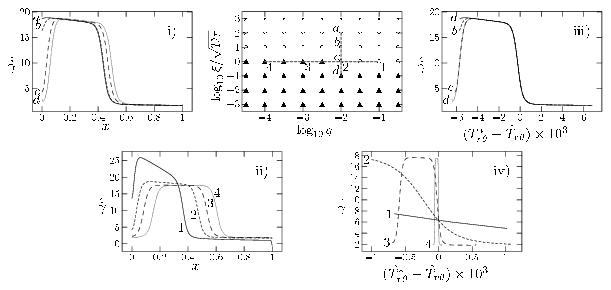}
  \end{center}
  \caption{ Flow profiles as a function of extrapolation length
    $\xi=\hat{\mathcal D}/\hat{W}$ and geometrical curvature $q$ for
    the DRP model with mixed boundary conditions, for
    $\langle\hat{\dot{\gamma}}\rangle = 3.8, \hat{\mathcal
      D}=7\times10^{-4}$ and $\ten{\Sigma}_0$ equal to the
    viscoelastic stress in the low shear rate branch,
    $\ten{\Sigma}_L$.  (i,iii) Shear rate $\dot{\gamma}$ as a function
    of position $x$; the labels $a,\ldots,d$ and $1,\ldots,4$ refer to
    the points in the $(\xi,q)$ parameter map in the center. (ii, iv):
    Shear rate as a function of the deviation of the total stress
    $T_{r\theta}\sim e^{-2qx}$ from the selected stress $T^{\ast}$.
    (Center) Map of different effective boundary conditions as a
    function of extrapolation length $\xi$ and curvature $q$:
    $\star\simeq$ Neumann boundary conditions; $\blacktriangle\simeq$
    Dirichlet conditions; $\circ\simeq$ mixed boundary conditions.}
\label{fig:RPvals}
\end{figure*}
Figure \ref{fig:RPvals} shows the results for the DRP model for
boundary conditions $\ten{\Sigma}_0=\ten{\Sigma}_L$. Consider changing
the anchoring strength at fixed $q$ (profiles \{i,ii,iii,iv\} in
Fig.~\ref{fig:RPvals}). For weak anchoring strength (large $\xi$,
profile a), the effective boundary condition at the inner wall is
nearly zero gradient (Neumann), and the high shear rate band lies near
the inner wall. Upon increasing the anchoring strength (reducing
$\xi$) the boundary conditions deviates slightly from Neumann
conditions before reverting to Dirichlet conditions for strong
enough anchoring (small $\xi$), thus imposing $\ten{\Sigma}_L$ at the
wall and inducing three bands. The crossover occurs when the
extrapolation length is of order the interfacial thickness,
$\xi\simeq\ell=\sqrt{D\tau}$.

In all cases the interface closest to the outer cylinder (at larger
$x$) remains at the selected stress $T^{\ast}_{r\theta}$ while the
inner interface is at a higher stress, as can be seen by the
intersections of the shear rate profiles as a function of total stress
in Fig.~\ref{fig:RPvals}(iii,iv). Upon approaching the flat limit both
interfaces approach the selected stress.  The degree of anchoring
strength above which the boundary conditions become Dirichlet-like
depends weakly on the geometric curvature. For a more highly curved
geometry a larger anchoring potential (smaller $\xi$) is required to
enforce the boundary condition $\ten{\Sigma}_0=\ten{\Sigma}_L$.  This
is because the higher stress at the inner wall competes with the
tendency of the wall to induce the low shear rate band.

\begin{figure*}[!htb]
\begin{center}
\subfigure[DJS model with mixed boundary conditions.]{
\begin{tabular}{c}
\includegraphics[width = 0.9\textwidth]{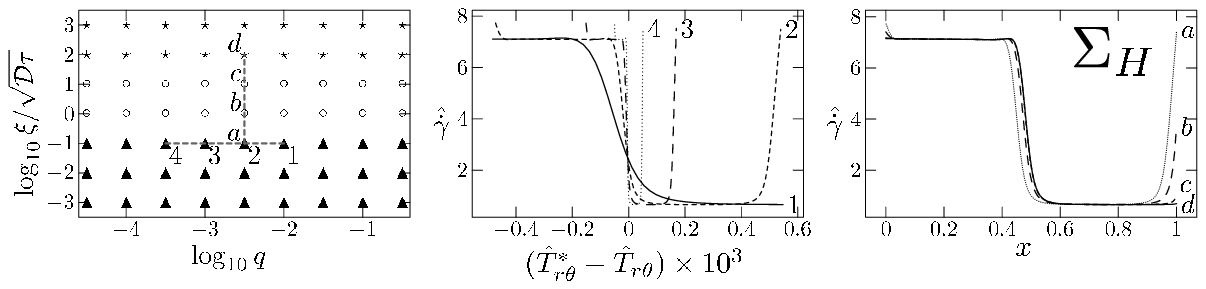}\\
\includegraphics[width = 0.9\textwidth]{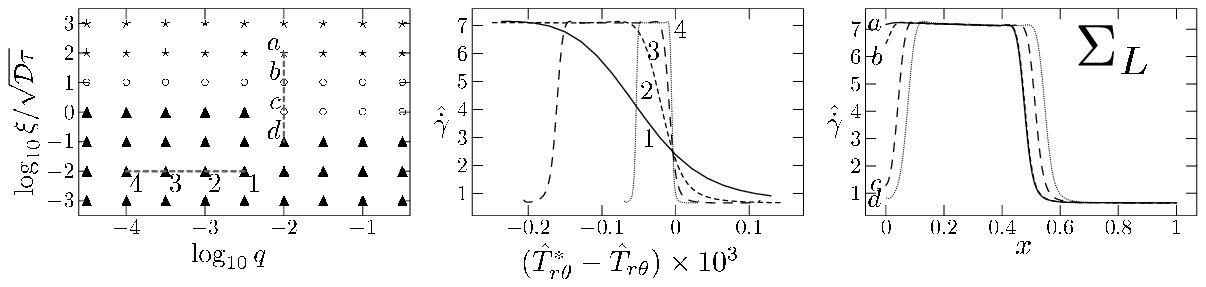}
\end{tabular}
\label{fig:mixedJS}
}
\subfigure[DRP model with mixed boundary conditions.]{ 
\begin{tabular}{c}
\includegraphics[width = 0.9\textwidth]{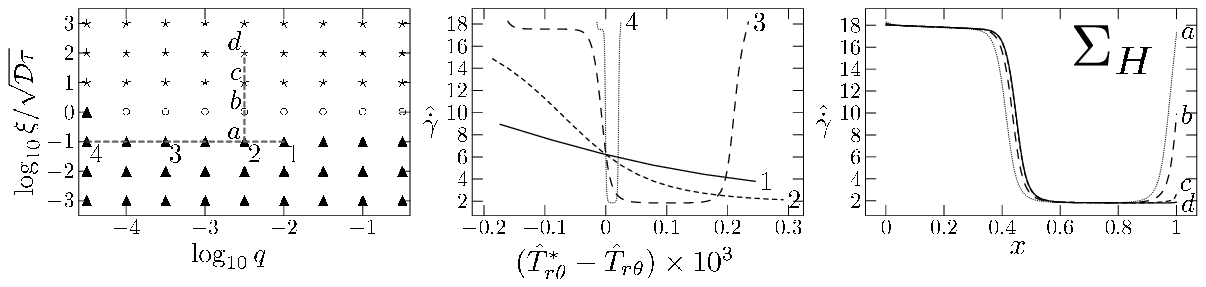}
\\
\includegraphics[width = 0.9\textwidth]{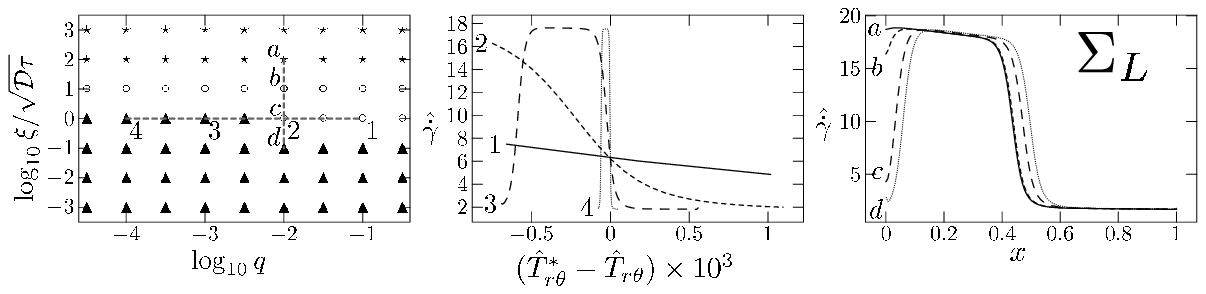}
\end{tabular}
\label{fig:mixedRP}
}
\end{center}
\caption{Left: Map of effective boundary conditions as a function of
  curvature $q$ and extrapolation length $\xi$ and selected shear rate
  profiles as a function of position and total stress, for
  $\mathcal{D}=7 \times 10^{-4}$: $\star\simeq$ Neumann boundary
  conditions; $\blacktriangle\simeq$ Dirichlet conditions;
  $\circ\simeq$ intermediate boundary conditions.  Shear rate as a
  function of local total stress (middle) and position (right) for the
  DJS and DRP models with boundary conditions
  $\ten{\Sigma}_0=\ten{\Sigma}_H$ and $\ten{\Sigma}_0=\ten{\Sigma}_L$.
}
\label{fig:mixedall}
\end{figure*}

We have examined whether or not this ``surface transition'' between
different anchoring states has an appreciable mechanical signal. The
torque at the inner cylinder, as would be measured in an experiment,
changes by of order $1\%$ upon tuning between Neumann-like and
Dirichlet-like boundary conditions; hence this has only a small effect
on the macroscopic response of the system. The degree of anchoring is
only weakly affected by the overall average shear rate; only when one
interface approaches the wall is there an effect, and again this is
reflected in only a small change in the measured total stress.

Similar behavior is found for $\ten{\Sigma}_0=\ten{\Sigma}_H$, with a
few differences.  In this case the interface closest to the inner wall lies
at the selected stress, as can be seen in the profiles for
$\dot{\gamma}(T_{r\theta})$ in Fig.~\ref{fig:mixedall}; this implies
that for both $\ten{\Sigma}_0=\ten{\Sigma}_L$ and
$\ten{\Sigma}_0=\ten{\Sigma}_H$ the interface that separates an inner
band at high shear rate from an outer band at low shear rate lies at
the selected stress.  Another difference is that the geometric
curvature $q$ has a very weak effect on the effective boundary
condition,  because the higher stress at the inner wall doesn't
compete with the boundary condition.

\section{Conclusion}
\label{sec:conclusion}
\subsection{Summary of results}
\label{sec:mainresults}
We have performed a numerical study of the effect of the boundary
conditions on the viscoelastic (polymer) stress for the diffusive
Johnson-Segalman (DJS) and diffusive non-stretching Rolie-Poly (DRP)
models, in cylindrical Couette flow. The main results are the
following:
\begin{enumerate}
\item[a.] Dirichlet boundary conditions (fixed $\ten{\Sigma}_0$) influence
  the possible hysteretic behavior for the flow curve: for
  $\ten{\Sigma}_0$ resembling the high shear rate phase a hysteresis
  loop occurs at the high shear rate end of the stress plateau, while
  for $\ten{\Sigma}_0$ resembling the low shear rate phase a
  hysteresis loop occurs at the low shear rate end of the stress
  plateau.
\item[b.] The walls can induce a lubrication (or thickened) layer that
  suppresses the hysteresis that would occur upon sweeping the shear
  rate into a banded state from a flow branch with characteristic
  dissimilar to the boundary layer.
\item[c.] As with Neumann boundary conditions, the hysteresis loop
  shrinks with decreasing diffusion constant. 
\item[d.] For Dirichlet boundary conditions the stress distribution in the
  gap depends on both the boundary condition and the stress gradient.
  A three-band state is stable for weak curvature such as cone and
  plate flow, which becomes more asymmetric or even two-banded as the
  stress gradient increases to that of typical cylindrical Couette
  flow.
\item[e.] For mixed boundary conditions the strength of
  wall anchoring determines whether the effective boundary conditions
  are Neumann or Dirichlet. The crossover occurs when the
  extrapolation $\xi$ length is of order the interfacial width $\ell$,
  or $W\simeq\sqrt{D\tau}$.
\end{enumerate}

This is the first use of the DRP model to study shear banding. The
model is microscopically motivated by either polymer solutions or
wormlike micelles, and its non-monotonic behavior arises from
better-understood physics (convected constraint release competing with
tube alignment) than that in the DJS model (where the ``slip
parameter'' $a$ is necessary). The physics of convected constraint
release yields a less well aligned state in the high shear rate phase,
whose alignment angle, compared to that from the DJS model, is closer
to that seen experimentally \cite{lerouge00}.  Numerically, interfaces
typically travel faster in the DRP model than in the DJS model, and
hence can be more quickly and reliably calculated, and the DRP model
is more robust to large time steps and inhomogeneous initial
conditions.

\subsection{Discussion and outlook}
\label{sec:discussion}

An important open problem is the relation of wall slip to shear
banding flows.  A boundary condition such as
$\ten{\Sigma}_0\sim\mathbf{\hat{v}}\mathbf{\hat{v}}$, which imposes
alignment parallel to the flow direction $\vec{v}$, yields velocity
profiles that could be interpreted in terms of wall slip, because of
the lack of a shear component of the viscoelastic stress at the wall.
The banding profiles in Fig.~\ref{fig:HdistSBJS}, for example, have
small regions of very high shear rate near either wall. Becu
\textit{et al.} measured a correlation between band motion and wall
slip in a solution of CTAB wormlike micelles
\cite{Becu.Manneville.ea04,becu2007etd}. They used smooth and
sand-blasted Couette cells and found different velocity profiles and
interface dynamics in the banding regime. It is possible that these
differences are a combination of both different intrinsic boundary
conditions and different degrees of wall slip.

 Another signature of
this effect would lie in the measurement of the low shear rate branch
before shear banding occurs: the flow cell (for example, the smooth
one) that induced tangential ordering would have a smaller stress at a
given shear rate, due to the associated lubrication layer, than would
the flow cell (\textit{e.g.} the sand-blasted one) with less
preference for the high shear rate branch.

In related work, Manneville and co-workers \cite{manneville2007wss}
measured the rheology and velocity profiles in a triblock copolymer
solution that forms wormlike micelles, using heterodyne light
scattering. At $T=37^{\circ}$ the solutions did not shear band but
did exhibit wall slip, within the $\simeq40\,\mu\textrm{m}$ resolution
of the technique. At a higher temperature $T=39.4^{\circ}$, believed
to correspond to longer micelles, the solution exhibited a three-band
shear banding scenario with the high shear rate bands near the inner
wall. These data are consistent with the walls inducing preferential
ordering parallel to the walls with increasing temperature and longer
micelles, so that that wall slip could be either ``true'' slip or an
apparent slip due to the specific nature of the boundary conditions.
Other mechanisms for slip, such as micelles detaching from the wall
and disentangling near the surface, may also be relevant in this case
\cite{BroGen1992L,MigHer1993PRL,BlaGra1996PRL}.

In principle, mixed boundary conditions allows for multiple stable
inhomogeneous solutions, which could be susceptible to non-linear
perturbations such as local flow inhomogeneities, thermal
fluctuations, asperities, motor noise, \textit{etc}. The solutions
found here switched between effective Neumann and Dirichlet boundary
conditions as a function of anchoring strength, but easily externally
controllable quantities such as flow geometry and imposed average
shear rate had very weak effects, and the total stress differences
between the two kinds of anchoring was small (typically less than a
percent).

One could, in principle, envision more complex surface anchoring
behavior, such as a multi-welled wall potential that would govern a
surface phase transition between two different wall orientations
$\ten{\Sigma}_{0a}$ and $\ten{\Sigma}_{0b}$. The order parameter in
the bulk could then, in principle, drive the boundary condition
between the two potential wells, which would then have more dramatic
consequences.  Such a bulk-driven wall transition could also be
triggered by fluctuations, would then lead to erratic band motion
coupled to an apparent wall slip, depending on the details of the
specific measurements.

Another physical effect that has not been considered here is the
concentration degree of freedom, which in some cases probably leads to
a depletion layer (or in rarer cases perhaps coagulation) layer near
the surface. Ref.~\cite{rossimckinley06} included concentration in
their treatment, but without any specific wall-concentration coupling
aside from a zero flux condition. This is a promising direction for
future study.

\section{Acknowledgements}
We thank the Royal Commission of 1851 and the EPSRC (SMF,
GR/S29560/02) for support.
\appendix
\section{Viscoelastic stress boundary
  conditions}\label{sec:liqu-cryst-anal}
The dynamics of the DJS or DRP model are formulated in terms of the
viscoelastic (or polymeric) contribution to the total stress. The free
energy of a polymer solution is more commonly expressed in terms of
the non-linear deformation tensor $\ten{\Lambda}$. A simple free
energy functional that includes surface, bulk, and distortion elastic
terms is
\begin{align}
 F &= \frac12\int d^3r\left[ G \ten{\Lambda}^2 + K \left(\nabla
     \ten{\Lambda}\right)^2\right] + \frac12\int_Sd^2r\, {W_0}
     (\ten{\Lambda}-\ten{\Lambda}_0)^2, 
 \end{align}
 where $K$ is analogous to the Frank elastic terms that penalize
 distortions of order parameter in nematic liquid crystals, and $G$ is
 the modulus. For ordinary polymer (rubber) elasticity the deformation
 tensor is
 $\ten{\Lambda}\simeq\langle\vec{R}\vec{R}\rangle/(Nb^2)-\tfrac13\ten{I}$
 , where $N$ is the number of Kuhn steps per strand, $b$ is the Kuhn
 length, and $\vec{R}$ is the end-to-end distance of the strand, and
 the modulus is thus given by $G\simeq 3 k_{\textrm{B}}T c_s$, where
 $c_s$ is the number of strands per unit volume. A simple dynamics and
 boundary conditions for such a free energy (based on the analogy with
 liquid crystals) is
\begin{align}
  \partial_t \ten{\Lambda} &= -\frac{1}{\zeta}\left[ G \ten{\Lambda}
 - K\nabla^2\ten{\Lambda}\right] \\ 
0 &= {W_0}\left(\ten{\Lambda} -
 \ten{\Lambda}_0\right) + K\nabla^2 \ten{\Lambda}.
 \end{align}

 Since the viscoelastic stress $\ten{\Sigma}$ is related to the strain
 by a modulus, $\ten{\Sigma} = G \ten{\Lambda}$, we can rewrite the
 equations above as
 \begin{widetext}
   \begin{align}
     F &= \frac12\int d^3r\left[ \frac{1}{G} \ten{\Sigma}^2 +
       \frac{K}{G^2} \left(\nabla \ten{\Sigma}\right)^2\right] +
     \frac12\int_Sd^2r \frac{W_0}{G^2}
     (\ten{\Sigma}-\ten{\Sigma}_0)^2\label{eq:1}\\
     \partial_t \ten{\Sigma} &= -\frac{G}{\zeta}\left[ \ten{\Sigma} -
       \frac{K}{G}\nabla^2\ten{\Sigma}\right] + \ldots \label{eq:3}\\
     0 &= {W_0}\left(\ten{\Sigma} - \ten{\Sigma}_0\right) +
     {K}\boldsymbol{\hat{\textbf{n}}}\cdot\vec{\nabla} \ten{\Sigma},
     \label{eq:4}
   \end{align}
 \end{widetext}
where $\boldsymbol{\hat{\textbf{n}}}$ is the outward unit vector
normal to the surface. 
On the other hand, the dynamics of the DJS model is
\begin{align}
  \partial_t \ten{\Sigma} &= -\frac{1}{\tau} \left[\ten{\Sigma} +
    {\mathcal D}\tau\nabla^2\ten{\Sigma}\right] + \ldots,
\end{align}
from which we identify $\tau=\zeta/G$ and ${\mathcal D}\tau=K/G$. Thus,
the boundary condition for the DJS model, which follows from this
correspondence and from Eq.~(\ref{eq:4}), is
\begin{align}
  0 &= \frac{W_0}{G}\left(\ten{\Sigma} - \ten{\Sigma}_0\right) +
  {\mathcal D}\tau 
  \nabla \ten{\Sigma}.
 \end{align}
Here $W_0$ is the surface anchoring term that penalizes the strain
variable at the surface according to Eq.~(\ref{eq:1}). In the main
text we have used $W\equiv W_0/G$.

\section{ Estimates of $W$ and $\mathcal D$}
\label{sec:diffusionconstant}

One mechanism for such a potential is the steric exclusion of micelles
at the wall, which would favor an oblate deformation $\ten{\Lambda}$
and hence an oblate viscoelastic stress.  Such an effect was found in
Monte Carlo simulations of polymer melts, which showed a decrease in
the radius of gyration $R_g^{\perp}$ normal to the wall of $20\%$ and
a corresponding increase parallel to the wall, independent of the
attraction of chain segments to the wall \cite{bitsanis1990}.
Assuming weak perturbations due to the wall potential, the free energy
governing subsequent perturbations is governed by the entropic
elasticity of polymer chains: $ F/\textrm{chain}\simeq
\tfrac32{k_{\textrm{B}}T}\,
\textrm{Tr}\left(\ten{\Lambda}-\ten{\Lambda}_0\right)^2.$ Converting
to a wall potential according to
\begin{align}
  \frac{F_{\textrm{wall}}}{\textrm{Area}} &\simeq
  \frac{F}{\textrm{chain}}\,c_s\, 2 R_g \\
& = \frac{3k_{\textrm{B}}T c_s b}{\sqrt{6N}}
  \textrm{Tr}\left(\ten{\Lambda}-\ten{\Lambda}_0\right)^2,
\end{align}
we estimate $W_0\simeq k_{\textrm{B}} T c_s b \sqrt{6/N}$ (according to
Eq.~\ref{eq:1}), and $W\simeq b\sqrt{2 N/3}=2R_g$.
Here,  $c_s$ is the number of strands per unit volume.

The stress in a solution of semiflexible polymers, such as wormlike
micelles, has a contribution $\ten{\Sigma}$ from the tube orientation
$\langle\vec{u}\vec{u}\rangle$, as well as a liquid crystalline
contribution $\ten{\Sigma}_{LC}$.  Although liquid crystalline effects
are only appreciable near an isotropic-to-nematic transition, they are
nonetheless present. Hence, the total stress takes the form
\cite{hess76,olmsted90}
 \begin{equation}
   \label{eq:10}
   \ten{T} = \ten{\Sigma} + 2\eta\ten{D} + \frac{\delta
     F_{LC}}{\delta\ten{Q}} + \ldots,
 \end{equation}
where $F_{LC}$ is the liquid
crystalline contribution to the free energy and the neglected terms
are non-linear in both $\ten{Q}$ and $\ten{D}$.

For a semiflexible polymer solution $F_{LC}$ has been calculated by
Liu and Fredrickson, and depends on the persistence length $\ell_p$.
To lowest order in the nematic order parameter $\ten{Q}$ it is
\begin{equation}
  \label{eq:11}
  F_{LC} = \tfrac12\int_V\left[a \textrm{Tr}\left(\ten{Q}^2\right) +
    {\mathcal K}\,\partial_{\mu}\ten{Q}\boldsymbol{:}\partial_{\mu}\ten{Q}\right],
\end{equation}
where only a single Frank constant has been included here. This leads
to a liquid crystalline contribution to the stress of the form
\begin{equation}
  \label{eq:12}
  \ten{\Sigma}_{LC}=a\ten{Q} - {\mathcal K}\nabla^2\ten{Q}.
\end{equation}
In this case the parameters are \cite{liu93}
 \begin{align}
{\mathcal K}&=
\frac{45}{126\pi}\frac{k_{\textrm{B}}T\ell_p\phi}{D^2},&
 a&\simeq \frac{45 \phi k_{\textrm{B}}T}{\pi D^2\ell_p},\label{eq:liufred}
\end{align}
where $\phi$ is the volume fraction $\phi$ \cite{liu93} and $D$ is the
micellar diameter.  For wormlike micelles $\ell_p\simeq
20\,\textrm{nm}, D \simeq 2\,\textrm{nm}$ \cite{cates90}, leading to
${\mathcal K}\simeq0.37\,\phi\,k_{\textrm{B}}T/\textrm{nm}$.  Upon
converting the entire description to one in terms of the total
micellar stress $\ten{\Sigma}_{\textrm{tot}} = \ten{\Sigma} +
\ten{\Sigma}_{LC}$ and performing a gradient expansion, one finds the
following contribution to the stress diffusion coefficient due to
Frank-like elasticity:
\begin{equation}
{\mathcal D}\tau\simeq\frac{\mathcal
  K}{a}\simeq\frac{\ell_p^2}{126}.\label{eq:8}
\end{equation}
There may of course, be other non-local contributions, which could
depend on micellar size, concentration, hydrodynamics, or other
physics.


\end{document}